\documentclass[pra,twocolumn,showpacs]{revtex4}
\usepackage{amssymb,times}
\usepackage{amsfonts}
\usepackage{dcolumn}
\usepackage{amsmath}
\usepackage{graphicx}
\usepackage{pstricks}
\usepackage{amsmath}
\usepackage{amssymb}
\usepackage{subfigure}
\usepackage{braket}
\usepackage[normalem]{ulem}
\usepackage{comment}
\let\originaleqref\eqref
\renewcommand{\eqref}{Eq.~\originaleqref}
\def \be {\begin{equation}}
\def \ee {\end{equation}}

\def \tr {\operatorname{Tr}}
\begin{document}
\title{Compressed sensing quantum process tomography for superconducting
quantum gates}
\author{Andrey V. Rodionov$^1$, Andrzej Veitia$^1$, R. Barends$^2$, J. Kelly$^2$, Daniel Sank$^2$, J. Wenner$^2$, John M. Martinis$^2$, Robert L. Kosut$^3$, and Alexander N. Korotkov$^1$}
\affiliation{$^1$Department of Electrical Engineering, University of California, Riverside, California 92521, USA \\
$^2$Department of Physics, University of California, Santa Barbara,
California 93106, USA \\
$^3$SC Solutions, 1261 Oakmead Parkway, Sunnyvale, California  94085, USA
}

\date{\today}

\begin{abstract}
We apply the method of compressed sensing (CS) quantum process
tomography (QPT) to characterize quantum gates based on
superconducting Xmon and phase qubits. Using experimental data for a
two-qubit controlled-Z gate, we obtain an estimate for the process
matrix $\chi$ with reasonably high fidelity compared to full QPT,
but using a significantly reduced set of initial states and
measurement configurations. We show that the CS method still works
when the amount of used data is so small that the standard QPT would
have an underdetermined system of equations. We also apply the CS
method to the analysis of the three-qubit Toffoli gate with
numerically added noise, and similarly show that the method works
well for a substantially reduced set of data. For the CS calculations we
use two different bases in which the process matrix $\chi$ is approximately
sparse, and show that the resulting estimates of the process
matrices match each other with reasonably high fidelity. For both
two-qubit and three-qubit gates, we characterize the quantum process
by not only its process matrix and fidelity, but also by the
corresponding standard deviation, defined via variation of the state
fidelity for different initial states.
\end{abstract}


\maketitle

\section{Introduction}

An important challenge in quantum information science and quantum
computing is the experimental realization of high-fidelity quantum
operations on multi-qubit systems. Quantum process tomography
(QPT)~\cite{N-C, ChuangQPT1997, Poyatos-97} is a procedure devised
to fully characterize a quantum operation. The role of QPT in
experimental characterization of quantum gates is twofold. First, it
allows us to quantify the quality of the gate; that is, it tells us
how close the actual and desired quantum operations are. Second, QPT
may aid in diagnosing and correcting errors in the experimental
operation~\cite{Boulant-03,Bendersky-08,Mohseni-09,Kofman,
Korotkov-13}. The importance of QPT has led to extensive theoretical
research on this subject (e.g.,
\cite{Leung-03,DAriano-03,Emerson-07,Mohseni-06,Wolf-08,Bogdanov-10}).

 Although conceptually simple, QPT suffers from a
fundamental drawback: the number of required experimental
configurations scales exponentially with the number of qubits
(e.g.,~\cite{MohseniResourseAnalysis}). An $N$-qubit quantum
operation can be represented by a $4^{N}\times 4^{N}$ process matrix
$\chi$~\cite{N-C} containing $16^{N}$ independent real parameters
(or $16^{N}-4^{N}$ parameters for a trace-preserving operation)
which can be determined experimentally by QPT. Therefore, even for
few-qubit systems, QPT involves collecting large amounts of
tomographic data and heavy classical postprocessing. To alleviate
the problem of exponential scaling of QPT resources, alternative
methods have been developed, e.g.,  randomized
benchmarking~\cite{Knill2008, Emerson-05, Magesan-12} and Monte
Carlo process certification~\cite{FlammiaMonteCarlo2011,daSilva-11}.
These protocols, however, find only the fidelity of an operation
instead of its full process matrix. Both randomized benchmarking and
Monte Carlo process certification have been demonstrated
experimentally for superconducting qubit gates
(see~\cite{Chow-09,Corcoles2013,SteffenMonteCarlo2012} and
references therein). Although these protocols are efficient tools
for the verification of quantum gates, their limitation lies in the
fact that they do not provide any description of particular errors
affecting a given process and therefore they cannot be used to
improve the performance of the gates.

 Recently, a new
approach to QPT  which incorporates ideas from signal processing
theory has been proposed~\cite{KosutSVD, ShabaniKosut}. The basic
idea  is to combine standard QPT with compressed sensing (CS) theory
\cite{Candes2006, Candes2008, Donoho, CandesWakin}, which asserts
that sparse signals may be efficiently recovered even when heavily
undersampled. As a result, compressed sensing quantum process
tomography (CS QPT) enables one to recover the process matrix $\chi$
from far fewer experimental configurations than standard QPT. The
method proposed in~\cite{KosutSVD, ShabaniKosut} is hoped to provide
an exponential speed-up over standard QPT. In particular, for a
$d$-dimensional system the method is supposed to require only
$O(s\log{d})$ experimental probabilities to produce a good estimate
of the process matrix $\chi$, if $\chi$ is known to be
$s$-compressible~\cite{s-sparse} in some known basis. (For
comparison, standard QPT requires at least $d^4$ probabilities,
where $d=2^{N}$ for $N$ qubits.) Note that there are bases in which
the process matrix describing the target process (the desired
unitary operation) is maximally sparse, i.e.\ containing only one
non-zero element; for example, this is the case for the so-called
singular-value-decomposition (SVD) basis \cite{KosutSVD} and the
Pauli-error basis \cite{Korotkov-13}. Therefore, if the actual
process is close to the ideal (target) process, then it is plausible
to expect that its process matrix is approximately sparse when written in
such a basis \cite{ShabaniKosut}.
 The CS QPT
method was experimentally validated in Ref.~\cite{ShabaniKosut} for
a photonic two-qubit controlled-Z (CZ) gate.  In that experiment, sufficiently accurate
estimates for the process matrix were obtained via CS QPT
using much fewer experimental configurations than the
standard QPT.

The CS idea also inspired another (quite different) algorithm for
quantum state tomography (QST) \cite{GrossFlammia2010,Flammia2012},
which can be generalized to QPT \cite{Flammia2012,Baldwin-14}. This matrix-completion method of
CS QST estimates the density matrices of nearly pure (low rank~$r$)
$d$-dimensional quantum states from expectation values of only $O(r
d \, {\rm poly} \log d)$ observables, instead of $d^2$ observables required for
standard QST. It is important to mention that this method does not
require any assumption about the quantum state of a system, except
that it must be a low-rank state (in particular, we do not need to
know the state approximately). The CS QST method has been used to
reconstruct the quantum states of a 4-qubit photonic
system~\cite{Yuan} and cesium atomic spins ~\cite{SmithDeutsch}. In
Ref.~\cite{Flammia2012} it has been shown that using the
Jamio\l{}kowski process-state isomorphism ~\cite{Jamiolkowski} the formalism of CS
QST can also be applied to the QPT, requiring $O(r d^{2}\, {\rm poly}\log d)$
measured probabilities (where $r$ is the rank of the Jamio\l{}kowski
state) to produce a good estimate of the process matrix $\chi$.
Therefore there is crudely a square-root speedup compared with
standard QPT.
 Note that this algorithm requires exponentially more
resources than the CS QPT method of Ref.\ \cite{ShabaniKosut}, but
it does not require to know a particular basis in which the matrix
$\chi$ is sparse. The performance of these two methods has been compared in the recent paper \cite{Baldwin-14} for a simulated quantum system with dimension $d=5$; the reported result is that the method of Ref.\ \cite{Flammia2012} works better for coherent errors, while the method of Ref.\ \cite{ShabaniKosut} is better for incoherent errors.

In this paper we apply the method of Ref.\ \cite{ShabaniKosut} to
the two-qubit CZ gate realized with superconducting qubits. Using
the experimental results, we find that CS QPT works reasonably well
when the number of used experimental configurations is up to $\sim$7
times less than for standard QPT. Using simulations for a
three-qubit Toffoli gate, we find that the reduction factor is
$\sim$40, compared with standard QPT. In the analysis we calculate
two fidelities: the fidelity of the CS QPT-estimated process matrix
$\chi_{\rm CS}$ compared with the matrix $\chi_{\rm full}$ from the
full data set and compared with $\chi_{\rm ideal}$ for the ideal
unitary process. Besides calculating the fidelities, we also
calculate the standard deviation of the fidelity, defined via the
variation of the state fidelity for different initial states. We
show that this characteristic is also estimated reasonably well by
using the CS QPT.

Our paper is structured as follows. Section~\ref{SQPT} is a brief
review of standard QPT and CS QPT. In Sec.~\ref{Configurations} we
discuss the set of measurement configurations used to collect QPT
data for superconducting qubits, and also briefly discuss our way to
compute the process matrix $\chi$ via compressed sensing. In
Sec.~\ref{results-two-qubits} we present our numerical results for
the CS QPT of a superconducting two-qubit CZ gate. In this section
we also compare numerical results obtained by applying the CS QPT
method in two different operator bases, the Pauli-error basis and
the SVD basis. In Sec.~\ref{OurResults-ThreeQu} we study the CS QPT
of a simulated three-qubit Toffoli gate with numerically added
noise. Then in Sec.~\ref{FidSquaredSection} we use the process
matrices obtained via compressed sensing to estimate the standard
deviation of the state fidelity, with varying initial state. Section
\ref{Conclusion} is a conclusion. In Appendices we discuss the
Pauli-error basis (Appendix A), SVD basis (Appendix B), and
calculation of the average square of the state fidelity (Appendix
C).

\section{Methods of Quantum Process Tomography}
\label{SQPT}

\subsection{ Standard Quantum Process Tomography}
\label{QPT-standard}

The idea behind QPT is to reconstruct a quantum operation $\rho^{\rm
in}\mapsto \rho^{\rm fin}= \mathcal{E}(\rho^{\rm in})$ from
experimental data. The quantum operation is a
completely positive  map, which for an $N$-qubit system prepared in the state
with density matrix $\rho^{\rm in}$ can be written as
\begin{equation}
\label{MainDefinition}
   \mathcal{E}(\rho^{\rm in})=\sum_{\alpha,\beta=1}^{d^{2}}\chi_{\alpha \beta} E_{\alpha}\rho^{\rm in} E_{\beta}^{\dagger},
\end{equation}
where $d=2^{N}$ is the dimension of the system, $\chi \in
\mathbb{C}^{d^{2}\times d^{2}}$ is the process matrix and
$\{E_{\alpha} \in \mathbb{C}^{d\times d} \} $ is a chosen basis of
operators. We assume that this basis is orthogonal,
$\braket{E_{\alpha}|E_{\beta}} \equiv
\tr(E_{\alpha}^{\dagger}E_{\beta})=Q \, \delta_{\alpha \beta}$,
where $Q=d$ for the Pauli basis and Pauli-error basis, while $Q=1$
for the SVD basis (see Appendices A and B). Note that for a trace-preserving operation ${\rm Tr}(\chi)=1$ if $Q=d$, while ${\rm Tr}(\chi)=d$ if $Q=1$.
In this paper we
implicitly assume the usual normalization $Q=d$, unless mentioned otherwise. The process matrix
$\chi$ is positive semidefinite (which implies being Hermitian), and
we also assume it to be trace-preserving,
\begin{eqnarray}
\label{ChiPositive}
\chi \geq   0 \quad (\text{positive semidefinite}), \\
\label{ChiTracePreserving}
\sum_{\alpha, \beta=1}^{d^{2}} \chi_{\alpha \beta}E_{\beta}^{\dagger}E_{\alpha}= \mathbb{I}_{d} \quad (\text{trace preserving}).\\ \notag
\end{eqnarray}
 These conditions ensure that $\rho^{\rm fin}=\mathcal{E}(\rho^{\rm in})$ is a legitimate density matrix for an arbitrary (legitimate) input state
 $\rho^{\rm in}$. The condition (\ref{ChiTracePreserving}) reduces the
 number of real independent parameters in $\chi$ from $d^4$  to $d^{4}-d^{2}$. Hence, the number of parameters needed to fully specify the map
 $\mathcal{E}$ scales as $O(16^{N})$ with the number of qubits $N$. Note that the set of allowed process matrices $\chi$ defined by
 Eqs.~(\ref{ChiPositive})~and~(\ref{ChiTracePreserving}) is convex \cite{Kosut2004Convex,KosutSVD}.

The essential idea of standard QPT is to exploit the linearity of
the map (\ref{MainDefinition}) by preparing the qubits in different
initial states, applying the quantum gate, and then measuring a set
of observables until the collected data allows us to obtain the
process matrix $\chi$ through matrix inversion or other methods.
More precisely, if the qubits are prepared in the state $\rho^{\rm
in}_{k}$, then the probability of finding them in the (measured)
state $\ket{\phi_{i}}$ after applying the gate  is given by
    \begin{equation}
\label{Probability}
P_{ik}=\tr(\Pi_{i} \mathcal{E}(\rho^{\rm in}_{k}))=\sum_{\alpha, \beta}\tr( \Pi_{i} E_{\alpha}\rho^{\rm in}_{k} E^{\dagger}_{\beta}) \chi_{\alpha
\beta},
    \end{equation}
where $\Pi_{i}=\ket{\phi_{i}}\bra{\phi_{i}}$. By preparing the qubits in one of the linearly independent input states $\{\rho^{\rm in}_{1},
\ldots \rho^{\rm in}_{N_{\rm in}} \}$ and performing a series of  projective measurements $ \{\Pi_{1},\ldots, \Pi_{N_{\rm meas}}\}$ on the output states,
one obtains a set of $m= N_{\rm in}N_{\rm meas}$ probabilities $ \{ P_{ik}\} $ which, using Eq.~(\ref{Probability}), may be written as
    \begin{equation}
\label{Vectorized}
\vec{P}(\chi)=\Phi \vec{\chi},
    \end{equation}
where $\vec{P}(\chi) \in \mathbb{C}^{m\times 1}$ and $\vec{\chi} \in
\mathbb{C}^{d^{4}\times 1}$ are vectorized forms of  $\{P_{ik}\} $
and $ \chi_{\alpha \beta}$, respectively. The  $m\times d^{4}$
transformation matrix $\Phi$ has entries given by $\Phi_{ik, \alpha
\beta}=\tr( \Pi_{i}
E_{\alpha}\rho^{\rm in}_{k} E^{\dagger}_{\beta})$.\\

In principle, for tomographically complete sets of input states
$\{\rho^{\rm in}_{1}, \ldots \rho^{\rm in}_{N_{\rm in}}\}$ and
measurement operators $\{\Pi_{1},\ldots, \Pi_{N_{\rm meas}}\}$, one
could invert Eq.~(\ref{Vectorized}) and thus uniquely find  $\chi$
by using the experimental set of probabilities
$\vec{P}^{\textrm{exp}}$. In practice, however, because of
experimental uncertainties present in $\vec{P}^{\textrm{exp}}$,
the process matrix thus obtained may be non-physical, that is, inconsistent
with the conditions (\ref{ChiPositive}) and
(\ref{ChiTracePreserving}). In standard QPT this problem is remedied
by finding the {\it physical} process matrix [satisfying (\ref{ChiPositive}) and
(\ref{ChiTracePreserving})] that minimizes (in some
sense)  the  difference between the probabilities $\vec{P}(\chi)$
and the experimental
probabilities $\vec{P}^{\textrm{exp}}$.\\

Two popular methods used to estimate a physical process matrix
$\chi$ compatible with the experimental data are the maximum
likelihood (ML) method ~\cite{MLEIterative,RiebeMLEQPT,Micuda-14}
(see also \cite{James2001,Bogdanov-04}) and the least-squares (LS)
method ~\cite{OBrien,Chow-09,Fuchs-12}. The ML method minimizes the
cost function \cite{MLEIterative}
    \be
    {\cal C}_{ML}=- \sum\nolimits_j P^{\rm exp}_j \ln P_j(\chi),
    \label{cost-ML}\ee
where the index $j$ labels the measured probabilities, while the LS
method (often also called maximum likelihood) minimizes the
difference between $\vec{P}(\chi)$ and $\vec{P}^{\textrm{exp}}$ in
the $\ell_{2}$-norm sense~\cite{NormDef}, so the minimized cost
function is
    \be
    {\cal C}_{LS}= ||\vec{P}(\chi)-\vec{P}^{\rm exp}||_{\ell_{2}}^2=
    \sum\nolimits_j [P^{\rm exp}_j- P_j(\chi)]^2.
    \label{cost-LS}\ee
In both methods the conditions (\ref{ChiPositive}) and
(\ref{ChiTracePreserving}) should be satisfied to ensure that $\chi$
corresponds to a physical process. This can be done in a number of
ways, for example, using the Cholesky decomposition, or Lagrange
multipliers, or just stating the conditions (\ref{ChiPositive}) and
(\ref{ChiTracePreserving}) as a constraint (if an appropriate software package is
used). The ML method (\ref{cost-ML}) is natural when the inaccuracy of
$\vec{P}^{\rm exp}$ is dominated by the statistical error due to a
limited number of experimental runs. However, this method does not
work well if a target probability $P_j$ is near zero, but $P_j^{\rm
exp}$ is not near zero due to experimental imperfections (e.g.,
``dark counts''); this is because the cost function (\ref{cost-ML})
is very sensitive to changes in $P^{\rm exp}_j$ when
$P_j(\chi)\approx 0$. Therefore, the LS method (\ref{cost-LS}) is a
better choice when the inaccuracy of $\vec{P}^{\rm exp}$ is not due
to a limited number of experimental runs.

Note that other cost functions can also be used for minimization in
the procedure. For example, by replacing $\ln P_j(\chi)$ in Eq.\
(\ref{cost-ML}) with $\ln [P_j(\chi)/P^{\rm exp}_j]$ (this obviously
does not affect optimization), then expanding the logarithm to
second order, and using condition $\sum_j P_j(\chi)=\sum_j P^{\rm
exp}_j$ (which cancels the first-order term), we obtain
\cite{James2001} ${\cal C}_{ML}\approx {\rm const} + \sum_j [P^{\rm
exp}_j- P_j(\chi)]^2/2P^{\rm exp}_j$. This leads to another natural
cost function
    \be
    {\cal C}=\sum_j \frac{[P_j(\chi)-P^{\rm exp}_j]^2}{P^{\rm exp}_j+a},
    \label{cost-3}\ee
where we phenomenologically introduced an additional parameter $a$,
so that for $a\gg 1$ the minimization reduces to the LS method,
while for $a \ll 1$ it is close to the ML method (the parameter $a$
characterizes the relative importance of non-statistical and
statistical errors). One more natural cost function is similar to
Eq.\ (\ref{cost-3}), but with $P^{\rm exp}_j$ in the denominator
replaced by $P^{\rm exp}_j(1-P^{\rm exp}_j)$ (see
\cite{MLEIterative}), which corresponds to the binomial distribution
variance.

In this paper we use the LS method (\ref{cost-LS}) for the standard
QPT. In particular, we find the process matrix $\chi_{\rm full}$ for
the full data set $\vec{P}^{\text{exp}}_{\rm full}$ by minimizing
$||\vec{P}(\chi_{\rm full})-\vec{P}^{\text{exp}}_{\rm
full}||_{\ell_{2}}$, subject to conditions Eqs.~(\ref{ChiPositive})
and (\ref{ChiTracePreserving}). Note that such minimization is a
convex optimization problem and therefore computationally tractable.

\subsection{Compressed Sensing Quantum Process Tomography}
\label{CS-QPT}

If the number of available experimental probabilities is smaller
than the number of independent parameters in the process matrix
(i.e. $ m< d^{4}- d^{2}$), then  the set of linear equations
Eq.~(\ref{Vectorized}) for the process matrix $\chi$ becomes
underdetermined. Actually, the LS method may still formally work in
this case for some range of $m$, but, as discussed in Secs.\
\ref{Sec-2q-LS} and \ref{OurResults-ThreeQu}, it is less effective.

By using the ideas of compressed sensing \cite{Candes2006, Donoho,
Candes2008, CandesWakin}, the method of CS QPT  requires a
significantly smaller set of experimental data to produce a
reasonably accurate estimate of the process matrix. Let us formulate
the problem mathematically as follows: we wish to find the physical
process matrix $\vec{\chi}_{0}$ satisfying the equation
\begin{equation}
\label{CSProblem}
\vec{P}^{\textrm{exp}}=\Phi \vec{\chi}_{0}+ \vec{z},
\end{equation}
where the vector $\vec{P}^{\rm exp} \in \mathbb{C}^{m}$ (with $ m<
d^{4}-d^{2}$) and the matrix $\Phi \in \mathbb{C}^{m\times d^{4}}$
are given, while  $\vec{z}\in \mathbb{C}^{m}$ is an unknown noise
vector, whose elements are assumed to be bounded (in the
root-mean-square sense) by a known limit $\varepsilon$,
$||\vec{z}||_{\ell_{2}}/\sqrt{m} \leq \varepsilon$. While this
problem seems to be ill-posed since the available information is
both noisy and incomplete, in Ref.~\cite{Candes2006} it was shown
that if the vector $\chi_{0}$ is sufficiently sparse and the matrix
$\Phi$ satisfies the restricted isometry property (RIP), $\chi_{0}$
can be accurately estimated from Eq.~(\ref{CSProblem}). Note that
the CS techniques of Ref.~{\cite{Candes2006}} were developed in the
context of signal processing; to adapt \cite{KosutSVD}  these
techniques to QPT we also need to include the positivity and
trace-preservation conditions,   Eqs.\ (\ref{ChiPositive}) and
(\ref{ChiTracePreserving}).

The idea of CS QPT \cite{ShabaniKosut} is to minimize the
$\ell_1$-norm~\cite{NormDef} of $\vec{\chi}$  in a basis where
${\chi}$ is assumed to be approximately sparse. Mathematically, the method
is solving the following convex optimization problem:
    \begin{eqnarray}
\label{mainL1problem} && \text{minimize}\,\,\,
{||\vec{\chi}||}_{\ell_1} \, ,
    \\
\label{ConditionsL1Problem} && \text{subject to \,}{||\vec{P}(\chi)
-\vec{P}^{\rm exp}||}_{\ell_2}\bigl/\sqrt{m} \le \varepsilon \qquad \\
&& \hspace{1.6cm}  \text{and conditions
(\ref{ChiPositive}), (\ref{ChiTracePreserving}).} \nonumber
    \end{eqnarray}
As shown
in Refs.\ \cite{Candes2008, ShabaniKosut}, a faithful reconstruction
recovery of an approximately  $s$-sparse process matrix~$\chi_0$ via this
optimization is guaranteed (see below) if (i) the matrix~$\Phi$
satisfies the RIP condition,
\begin{equation}
\label{RIP} 1-\delta_s \le \displaystyle\frac{||\Phi\vec\chi_1 -
\Phi\vec\chi_2 ||^2_{\ell_2}}{||\vec\chi_1 -
\vec\chi_2||^2_{\ell_2}}\le 1+\delta_s ,
\end{equation}
for all $s$-sparse vectors (process matrices) $\vec{\chi}_1$ and
$\vec{\chi}_2$,
 (ii) the isometry constant $\delta_s$ is sufficiently small, $\delta_s < \sqrt{2} - 1$, and
 (iii) the number of data points is sufficiently large,
    \be
 m\ge C_0 s \log(d^4/s)=O(sN),
    \label{ineq-m}\ee
where $C_0$ is a constant.  Quantitatively, if $\chi_{\rm CS}$ is
the solution of the optimization problem [Eqs.~(\ref{mainL1problem})
and (\ref{ConditionsL1Problem})], then the estimation error
${||\chi_{\rm CS}-\chi_{0}||}_{\ell_{2}} $ is bounded as
 \begin{equation}
 \label{bounds}
 \frac{ || \chi_{\rm CS} - \chi_{0} ||_{\ell_{2}}}{\sqrt{m}} \leq
 \frac{ C_{1} ||\chi_{0}(s)-\chi_{0}||_{\ell_{1}} }{\sqrt{m s}}+C_{2\,} \varepsilon,
 \end{equation}
where $\chi_{0}(s)$ is the best $s$-sparse approximation of
$\chi_{0}$, while $C_{1}$ and $ C_{2}$ are constants of the order
$O(\delta_{s})$. Note that in the noiseless case ($\varepsilon=0 $)
the recovery is exact if the process matrix $\chi_0$ is $s$-sparse.
Also note that while the required number of data points $m$ and the
recovery accuracy depend on the sparsity $s$, the method itself
[Eqs.~(\ref{mainL1problem}) and (\ref{ConditionsL1Problem})] does
not depend on $s$, and therefore $s$ need not be known.

The inequality (\ref{ineq-m}) and the first term in the inequality
(\ref{bounds}) indicate that the CS QPT method is supposed to work
well only if the actual process matrix $\chi_0$ is sufficiently
sparse. Therefore, it is important to use an operator basis
$\{E_{\alpha}\}$ [see  Eq.~(\ref{MainDefinition})], in which the
ideal (desired) process matrix $\chi_{\rm ideal}$ is maximally
sparse, i.e., it contains only one nonzero element. Then it is
plausible to expect the actual process matrix $\chi_0$ to be approximately
sparse \cite{ShabaniKosut}. In this paper we will use two bases in
which the ideal process matrix is maximally sparse. These are the
so-called Pauli-error basis \cite{Korotkov-13} and the SVD basis of
the ideal unitary operation~\cite{KosutSVD}. In the Pauli-error
basis  $\{E_{\alpha}\}$, the first element $E_1$ coincides with the
desired unitary $U$, while other elements are related via the
$N$-qubit Pauli matrices $\cal P$, so that $E_\alpha=U{\cal
P}_\alpha$. In the SVD basis  $E_1=U /\sqrt{d}$, and other elements
are obtained via a numerical SVD procedure. More details about the
Pauli-error and SVD bases are discussed in Appendices  A and B.

As mentioned previously, the method of CS QPT involves the RIP condition
(\ref{RIP}) for the transformation matrix~$\Phi$. In Ref.\ \cite{ShabaniKosut}
it was shown that if the
transformation matrix~$\Phi$ in Eq.~(\ref{Vectorized}) is
constructed from randomly selected input states~$\rho_{k}^{\rm in}$
and random measurements~$\Pi_{i}$, then~$\Phi$ obeys the RIP
condition with high probability. Notice that once a basis
$\{E_{\alpha}\}$ and a  tomographically complete (or overcomplete)
set $\{\rho^{\rm in}_{k}, \Pi_{i}\}$ have been chosen, the matrix
$\Phi_{\textrm{full}}$ corresponding to the full data set is fully
defined, since it does not depend on the experimental outcomes.
Therefore, the mentioned above result of Ref.~\cite{ShabaniKosut}
tells us that if we build a matrix~$\Phi_{m}$ by randomly selecting
$m$ rows from~$\Phi_{\textrm{full}}$, then~$\Phi_{m}$ is very likely
to satisfy the RIP condition. Hence, the
submatrix~$\Phi_{m}~\in~\mathbb{C}^{m\times d^4}$, together with the
corresponding set of experimental outcomes
$\vec{P}^{\textrm{exp}}~\in~\mathbb{C}^{m}$ can be used to produce
an  estimate of the process matrix via the~$\ell_{1}$-minimization
procedure (\ref{mainL1problem}) and (\ref{ConditionsL1Problem}).

\section{Standard and CS QPT of multi-qubit superconducting gates}
\label{Configurations}

There are several different ways to perform standard QPT for an
$N$-qubit quantum gate realized with superconducting qubits
\cite{Mariantoni-11,Bialczak-10,Reed-12,Dewes-12,FedorovImplementToffoli,
Chow-12,Chow-13,Yamamoto-10}. The differences are the following.
First, it can be performed using either $n_{\rm in}=4$ initial
states for each qubit
\cite{Bialczak-10,Reed-12,Dewes-12,FedorovImplementToffoli},
 e.g.,  $\{\ket{0}, \ket{1},
(\ket{0}+\ket{1})/\sqrt{2}, (\ket{0}+i\ket{1})/\sqrt{2}\}$, or using
$n_{\rm in}=6$ initial states per qubit \cite{Chow-12,Chow-13},
$\{\ket{0}, \ket{1}, (\ket{0}\pm\ket{1})/\sqrt{2}, (\ket{0}\pm
i\ket{1})/\sqrt{2}\}$, so that the total number of initial states is
$N_{\rm in}=n_{\rm in}^N$. (It is tomographically sufficient to use
$n_{\rm in}=4$, but the set of 6 initial states is more symmetric,
so it can reduce the effect of experimental imperfections.) Second, the
final measurement of the qubits can be realized in the computational
basis after one out of $n_{\rm R}=3$ rotations per qubit
\cite{Bialczak-10,Dewes-12}, e.g., $\mathcal{R}_{\rm meas}=\{\mathbb{I},
R_{y}^{-\pi/2}, R_{x}^{\pi/2}\}$, or $n_{\rm R}=4$ rotations
\cite{Chow-09,Reed-12,Chow-13}, e.g., $\mathcal{R}_{\rm meas}=\{\mathbb{I}, R_{y}^{\pi}, R_{y}^{\pi/2}, R_{x}^{\pi/2}\}$, or $n_{\rm R}=6$ rotations
\cite{Mariantoni-11,Chow-12,Yamamoto-10}, e.g., $\mathcal{R}_{\rm meas}=\{\mathbb{I}, R_{y}^{\pi},
R_{y}^{\pm \pi/2}, R_{x}^{\pm\pi/2}\}$. This gives $N_{\rm R
}=n_{\rm R}^N$ measurement ``directions'' in the Hilbert space.
Third, it may be possible to measure the state of each qubit simultaneously
\cite{Bialczak-10,Mariantoni-11,Dewes-12}, so that the probabilities for all $2^N$ outcomes
are measured, or it may be technically possible to measure the probability for only one state (say, $|0... 0\rangle$) or a weighed sum of the probabilities  \cite{Chow-12,Reed-12,FedorovImplementToffoli}.
Therefore, the number of measured
probabilities for each configuration is either $N_{\rm prob}=2^N$ (with $2^N-1$
independent probabilities, since their sum is equal 1) or $N_{\rm
prob}=1$. Note that if $N_{\rm prob}=2^N$, then using $n_{\rm R}=6$
rotations per qubit formally gives the same probabilities as for
$n_{\rm R}=3$, and in an experiment this formal symmetry can be used
to improve the accuracy of the results. In contrast, in the case
when $N_{\rm prob}=1$, the use of $n_{\rm R}=4$ or $n_{\rm R}=6$ are
natural for the complete tomography.

Thus, the number of measurement configurations (including input
state and rotations) in standard QPT is $M_{\rm conf}=N_{\rm
in}N_{\rm R}=n_{\rm in}^N n_{\rm R}^N$, while the total number of
probabilities in the data set is $M=M_{\rm conf}N_{\rm prob}$. This
number of probabilities can be as large as $M=72^N$ for $n_{\rm
in}=6$, $n_{\rm R}=6$, and $N_{\rm prob}=2^N$ (with $72^N-36^N$
independent probabilities). Since only $16^N-4^N$ independent
probabilities are necessary for the standard QPT, a natural choice
for a shorter experiment is $n_{\rm in}=4$, $n_{\rm R}=3$, and
$N_{\rm prob}=2^N$; then $M=24^N$, with $24^N-12^N$ independent
probabilities. If $N_{\rm prob}=1$ due to the limitations of the
measurement technique, then the natural choices are $n_{\rm in}=4$
and  $n_{\rm R}=4$, giving $M=16^N$ or $n_{\rm in}=4$  and $n_{\rm
R}=6$, giving $M=24^N$.

In this paper we focus on the case $n_{\rm in}=4$, $n_{\rm R}=3$,
and $N_{\rm prob}=2^N$. Then for a two-qubit quantum gate there are
$M_{\rm conf}=12^N=144$ measurement configurations and $M=24^N=576$
probabilities (432 of them independent). For a three-qubit gate
there are $M_{\rm conf}=1728$ configurations and $M=13824$
probabilities (12096 of them independent).

The main experimental data used in this paper are for the two-qubit
CZ gate realized with Xmon qubits \cite{Xmon}. The data were
obtained with $n_{\rm in}=6$, $n_{\rm R}=6$, and $N_{\rm prob}=2^N$.
However, since the main emphasis of this paper is analysis of the
QPT with a reduced data set, we started by reducing the data set to
$n_{\rm in}=4$ and $n_{\rm R}=3$ by using only the corresponding
probabilities and removing other data. We will refer to these data
as ``full data'' (with $M_{\rm conf}=144$ and $M=24^N=576$). For
testing the CS method we randomly choose $m_{\rm conf} \leq M_{\rm
conf}$ configurations, with corresponding $m=4m_{\rm conf}$
experimental probabilities ($3m_{\rm conf}$ of them independent).
Since the process matrix $\chi$ is characterized by $16^N-4^N=240$
independent parameters, the power of the CS method is most evident
when $m_{\rm  conf}<80$, so that the system of equations
(\ref{Vectorized}) is underdetermined. [For a three-qubit gate the
system of equations becomes underdetermined for $m_{\rm
conf}<(16^N-4^N)/(2^N-1)=576$.]

The data used for the analysis here were taken on a different device
from the one used in Ref.\ \cite{Barends-14}. For the device used
here the qubits were coupled via a bus, and the entangling gate
between qubits A and B was implemented with three multiqubit
operations: 1) swap state from qubit B to bus, 2) CZ gate between
qubit A and bus, 3) swap back to qubit B. The swap was done with the
resonant Strauch gate \cite{Strauch-03}, by detuning the frequency
of qubit A with a square pulse. Generating a square pulse is
experimentally challenging, moreover this gate has a single optimum
in pulse amplitude and time. We also note that the qubit frequency
control was not optimized for imperfections in the control wiring,
as described in Ref.\ \cite{Kelly-14} (see also Fig.\ S4 in
Supplementary  Information of \cite{Barends-14}).
 The
combination of device, non-optimal control, and multiple operations,
leads to the experimental process fidelity $F_\chi = 0.91$ of the CZ
gate used for the analysis here to be significantly less than the
randomized benchmarking fidelity $F_{RB} = 0.994$ reported in
\cite{Barends-14}. Moreover, QPT necessarily includes state
preparation and measurement (SPAM) errors \cite{Magesan-12}, while
randomized benchmarking does not suffer from these errors. This is
why we intentionally used the data for a not-well-optimized CZ gate
so that the gate error dominates over the SPAM errors. (Note that we
use correction for the imperfect measurement fidelity
\cite{Mariantoni-11}; however, it does not remove the measurement
errors completely.) It should also be mentioned that in the ideal
case $1-F_\chi =(1-F_{\rm RB})\times (1+2^{-N})$, so the QPT
fidelity is supposed to be slightly less than the randomized
benchmarking fidelity.

For the full data set, we first calculate the process matrix
$\chi_{\rm full}$ by using the least-squares method described at the
end of Sec.\  \ref{QPT-standard}. For that we use three different
operator bases \{$E_{\alpha}$\}: the Pauli basis, the Pauli-error
basis, and the SVD basis. The pre-computed transformation matrix
$\Phi$ in Eq.\ (\ref{Vectorized}) depends on the choice of the
basis, thus giving a basis-dependent result for $\chi_{\rm full}$.
We then check that the results essentially coincide by converting
$\chi_{\rm full}$ between the bases and calculating the fidelity
between the corresponding matrices (the infidelity is found to be
less than $10^{-7}$). The fidelity between two process matrices
$\chi_{1}$ and $\chi_{2}$ is defined as the square of the Uhlmann
fidelity~\cite{Uhlmann,Jozsa},
\begin{equation}
\label{DefinitionFidelity} F(\chi_{1},\chi_{2}) =
\Bigl({\rm Tr}\sqrt{\chi_{1}^{1/2}\,\chi_{2}\,\chi_{1}^{1/2}}\Bigr)^2,
\end{equation}
so that it reduces to $F(\chi_{1},\chi_{2})=\tr(\chi_{1}\chi_{2})$
\cite{GilchristNielsen} when at least one of the process matrices
corresponds to a unitary operation. Since $0\leq F\leq 1$, we refer
to $1-F$ as the infidelity.

After calculating $\chi_{\rm full}$ for the full data set, we can calculate its fidelity compared to the process matrix $\chi_{\rm ideal}$ of the desired ideal unitary operation, $F_\chi = F_{\rm full}=F(\chi_{\rm full}, \chi_{\rm ideal})$. This is the main number used to characterize the quality of the quantum operation.

Then we calculate the compressed-sensing process matrix $\chi_{\rm
CS}$ by  solving the $\ell_{1}$-minimization problem described by
Eqs.\ (\ref{mainL1problem}) and (\ref{ConditionsL1Problem}), using
the reduced data set. It is obtained from the full data set by
randomly selecting $m_{\rm conf}$ configurations out of the full
number $M_{\rm conf}$ configurations. We use the fidelity
$F(\chi_{\rm CS},\chi_{\rm full})$ to quantify how well the process
matrix $\chi_{\rm CS}$ approximates the matrix $\chi_{\rm full}$
obtained from full tomographic data. Additionally, we calculate the
process fidelity $F(\chi_{\rm CS},\chi_{\rm ideal})$ between
$\chi_{\rm CS}$ and the ideal operation, to see how closely it
estimates the process fidelity $F_{\rm full}$, obtained using the
full data set.

Since both the least-squares  and the $\ell_{1}$-norm minimization
are convex optimization problems~\cite{KosutSVD,BoydConvex}, they
can be efficiently solved numerically. We used two ways for
MATLAB-based numerical calculations: (1) using the package CVX
\cite{CVXBoyd}, which calls the convex solver SeDuMi
\cite{SeDuMi13}; or (2) using the package YALMIP
\cite{YALMIPLofberg}, which calls the convex solver SDPT3
\cite{SDPT3}. In general, we have found that for our particular realization of computation, CVX with the solver SeDuMi works better than the
combination YALMIP-SDPT3 (more details are below).

\section{Results for two-qubit CZ gate}
\label{results-two-qubits}

In this section we present results for the experimental CZ gate
realized with superconducting  Xmon qubits \cite{Xmon,Barends-14}.
As explained above, the full data set consists of $M=576$ measured
probabilities (432 of them independent), corresponding to $M_{\rm
conf}=4^2\times 3^2=144$ configurations, with 4 probabilities (3 of
them independent) for each configurations. The LS method using the
full data set produces the process matrix $\chi_{\rm full}$, which
has the process fidelity $F(\chi_{\rm full}, \chi_{\rm
ideal})=0.907$ relative to the ideal CZ operation. Note that our
full data set is actually a subset of an even larger data set (as
explained in the previous section), and the $\chi$ matrix calculated
from the initial set corresponds to the process fidelity of 0.928;
the difference gives a crude estimate of the overall accuracy of the
procedure.

The CS method calculations were mainly done in the Pauli-error
basis, using the CVX-SeDuMi combination for the $\ell_1$-norm
minimization. This is what is implicitly assumed in this section,
unless specified otherwise. Note that the CS-method optimization is
very different from the LS method. Therefore, even for the full data
set we would expect the process matrix $\chi_{\rm CS}$ to be
different from $\chi_{\rm full}$. Moreover, $\chi_{\rm CS}$ depends
on the noise parameter $\varepsilon$ [see Eq.\
(\ref{ConditionsL1Problem})], which to some extent is arbitrary. To
clarify the role of the parameter $\varepsilon$, we will first
discuss the CS method applied to the full data set, with varying
$\varepsilon$, and then discuss the CS QPT for a reduced data set,
using either near-optimal or non-optimal values of $\varepsilon$.

\subsection{Full data set, varying $\varepsilon$ }
\label{varyingeps}

\begin{figure}[tb]
\includegraphics[width=8.5cm]{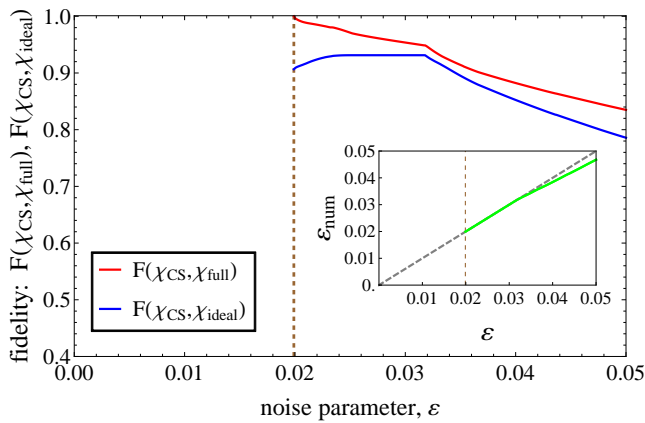}
\caption{(color online) The CS QPT procedure, applied to the full
data set, with varying noise parameter $\varepsilon$. The red
(upper) line shows the fidelity  $F(\chi_{\rm CS},\chi_{\rm full})$
between the process matrix $\chi_{\rm CS}$ obtained using the
compressed-sensing method and the matrix $\chi_{\rm full}$ obtained
using the least-squares method. The blue (lower) line shows the
process fidelity $F(\chi_{\rm CS},\chi_{\rm ideal})$, i.e., compared
with the matrix $\chi_{\rm ideal}$ of the ideal unitary process. The
vertical dashed brown line corresponds to the noise level
$\varepsilon_{\rm opt}=||\vec{P}^{\rm exp}_{\rm full}-\Phi
\vec{\chi}_{\rm full}||_{\ell_2}/\sqrt{M}=0.0199$ obtained in the LS
procedure. The inset shows $\varepsilon_{\rm num}=||\vec{P}^{\rm
exp}_{\rm full}-\Phi \vec{\chi}_{\rm CS}||_{\ell_2}/\sqrt{M}$ as a
function of $\varepsilon$ (green line); for comparison, the dashed
line shows the expected straight line, $\varepsilon_{\rm
num}=\varepsilon$.
  The numerical calculations have been carried out in the Pauli-error basis using CVX-SeDuMi package.}
 \label{fig1}
 \end{figure}

We start with calculating the process matrix $\chi_{CS}$ by solving
the $\ell_{1}$-minimization problem, Eqs.~(\ref{mainL1problem}) and
(\ref{ConditionsL1Problem}), using the full data set and varying the
noise parameter $\varepsilon$. The resulting matrix is compared with
the LS result $\chi_{\rm full}$ and with the ideal matrix $\chi_{\rm
ideal}$. Figure \ref{fig1} shows the corresponding fidelities
$F(\chi_{\rm CS}, \chi_{\rm full})$ and $F(\chi_{\rm CS}, \chi_{\rm
ideal})$ as functions of $\varepsilon$. We see that $\chi_{\rm CS}$
coincides with $\chi_{\rm full}$ [so that $F(\chi_{\rm CS},
\chi_{\rm full})=1$] at the optimal value $\varepsilon_{\rm
opt}=0.0199$. This is exactly the noise level corresponding to the
LS procedure,  $||\vec{P}^{\rm exp}_{\rm full}-\Phi \vec{\chi}_{\rm
full}||_{\ell_2}/\sqrt{M}=0.0199$. With $\varepsilon$ increasing above
this level, the relative fidelity between $\chi_{\rm CS}$ and
$\chi_{\rm full}$ decreases, but it remains above 0.95 for
$\varepsilon <0.028$. Correspondingly, the process fidelity reported
by $\chi_{\rm CS}$, i.e.\ $F(\chi_{\rm CS}, \chi_{\rm ideal})$, also
changes. It starts with $F(\chi_{\rm CS}, \chi_{\rm
ideal})=F(\chi_{\rm full}, \chi_{\rm ideal})=0.907$ for
$\varepsilon=0.0199$, then increases with increasing $\varepsilon$,
then remains flat above $\varepsilon =0.025$, and then decreases at
$\varepsilon> 0.032$. We note that for another set of experimental
data (for a CZ gate realized with phase qubits) there was no
increasing part of this curve, and the dependence of $F(\chi_{\rm
CS}, \chi_{\rm ideal})$ on $\varepsilon$ remained practically flat
for a wide range of $\varepsilon$; one more set of experimental data
for phase qubits again had the increasing part of this curve.

To check how close the result of $\ell_1$-optimization
(\ref{mainL1problem}) is to the upper bound of the condition
(\ref{ConditionsL1Problem}), we calculate the numerical value
$\varepsilon_{\rm num}=||\vec{P}^{\rm exp}_{\rm full}-\Phi
\vec{\chi}_{\rm CS}||_{\ell_2}/\sqrt{M}$ as a function of
$\varepsilon$. The result is shown in the inset of Fig.\ \ref{fig1},
we see that $\varepsilon_{\rm num}$ is quite close to $\varepsilon$.
The CVX-SeDuMi package does not solve the optimization problem for
values of the noise parameter $\varepsilon$ below the optimal value
$\varepsilon_{\rm opt}$.

Finding a proper value of $\varepsilon$ to be used in the CS method
is not a trivial problem, since for the reduced data set we
cannot find $\varepsilon_{\rm opt}$ in the way we used. Therefore,
the value of $\varepsilon$ should be estimated either from some
prior information about the noise level in the system or by trying
to solve the $\ell_1$-minimization problem with varying value of
$\varepsilon$. Note that the noise level $||\vec{P}^{\rm exp}-\Phi
\vec{\chi}_{\rm ideal}||_{\ell_2}/\sqrt{M}$ defined by the ideal
process is not a good estimate of $\varepsilon_{\rm opt}$; in
particular for our full data it is 0.035, which is significantly
higher than $\varepsilon_{\rm opt}=0.0199$.

\subsection{Reduced data set, near-optimal $\varepsilon$ }

\begin{figure}[tb]
\includegraphics[width=8.5cm]{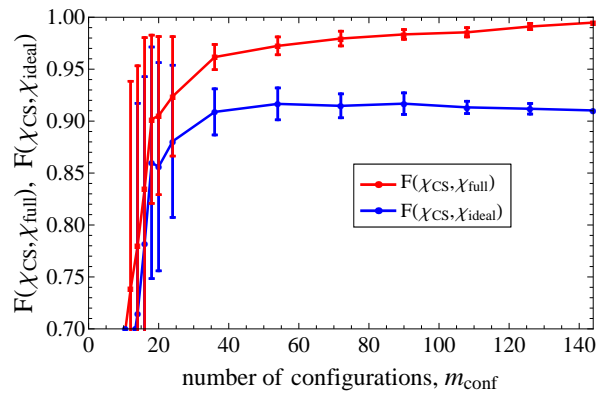}
\caption{(color online) The CS method results using a reduced data
set with randomly chosen $m_{\rm conf}$ configurations. The red
(upper) line shows the fidelity $F(\chi_{\rm CS},\chi_{\rm full})$
between the CS-estimated process matrix $\chi_{\rm CS}$ and the
matrix $\chi_{\rm full}$ obtained from the full data set. The blue
(lower) line shows the estimated process fidelity
$F_\chi=F(\chi_{\rm CS},\chi_{\rm ideal})$. The procedure of
randomly choosing $m_{\rm conf}$ out of 144 configurations is
repeated 50 times; the error bars show the calculated standard
deviations. The noise parameter $\varepsilon=0.002015$ is chosen
slightly above $\varepsilon_{\rm opt}$.  The calculations are
carried out in the Pauli-error  basis using CVX-SeDuMi. The
experimental data are for the CZ gate realized with Xmon qubits; the
process fidelity is $F(\chi_{\rm full},\chi_{\rm ideal})=0.907$.}
 \label{fig2}
\end{figure}

Now we apply the CS method to a reduced data set, by randomly
choosing $m_{\rm conf}$ out of $M_{\rm conf}=144$ configurations,
while using all 4 probabilities for each configuration. (Therefore
the number of used probabilities is $m=4m_{\rm conf}$ instead of
$M=4M_{\rm conf}$ in the full data set.) For the noise level
$\varepsilon$ we use a value slightly larger than $\varepsilon_{\rm
opt}$ \cite{ShabaniKosut}. If a value too close to $\varepsilon_{\rm
opt}$ is used, then the optimization procedure often does not find a
solution; this happens when we choose configurations with a
relatively large level of noise in the measured probability values.
For the figures presented in this subsection we used
$\varepsilon=0.02015$, which for the full data set corresponds to
the fidelity of 0.995 compared with $\chi_{\rm full}$ and to the process fidelity of 0.910 (see Fig.\ \ref{fig1}).

Figure \ref{fig2} shows the fidelities $F(\chi_{\rm CS}, \chi_{\rm
full})$ (upper line) and $F(\chi_{\rm CS}, \chi_{\rm ideal})$ (lower
line) versus the number $m_{\rm conf}$ of used configurations. For
each value of $m_{\rm conf}$ we repeat the procedure 50 times,
choosing different random configurations. The error bars in Fig.\
\ref{fig2} show the standard deviations ($\pm \sigma$) calculated
using these 50 numerical experiments, while the central points
correspond to the average values.

We see that the upper (red) line starts with fidelity $F(\chi_{\rm CS},
\chi_{\rm full})=0.995$ for the full data set ($m_{\rm conf}=144$) and decreases with decreasing $m_{\rm conf}$. It is important that this decrease is not very strong, so that we can reconstruct the process matrix reasonably accurately, using only a small fraction of the QPT data. We emphasize that the system of equations (\ref{Vectorized}) in the standard QPT procedure becomes underdetermined at $m_{\rm conf}<80$; nevertheless, the CS method reconstructs $\chi_{\rm full}$ quite well for $m_{\rm conf}\agt 40$ and still gives reasonable results for $m_{\rm conf}\agt 20$. In particular, for $m_{\rm conf}$ between 40 and 80, the reconstruction fidelity $F(\chi_{\rm CS},
\chi_{\rm full})$ changes between 0.96 and 0.98.

The lower (blue) line in Fig.\ \ref{fig2} shows that the process fidelity $F_{\chi}=F(\chi_{\rm CS},\chi_{\rm ideal})$ can also be found quite accurately, using  only $m_{\rm conf}\agt 40$ configurations (the line remains practically flat), and the CS method still works reasonably well down to  $m_{\rm conf}\agt 20$. Even though the blue line remains practically flat down to $m_{\rm conf}\simeq 40$, the error bars grow, which means that in a particular experiment with substantially reduced set of QPT data, the estimated process fidelity $F_\chi$ may noticeably differ from the actual value. It is interesting that the error bars become very large at approximately the same value ($m_{\rm conf}\simeq 20$), for which the average values for the red and blue lines become unacceptably low.

\begin{figure}[tb]
 \includegraphics[width=4.15cm,angle=-00]{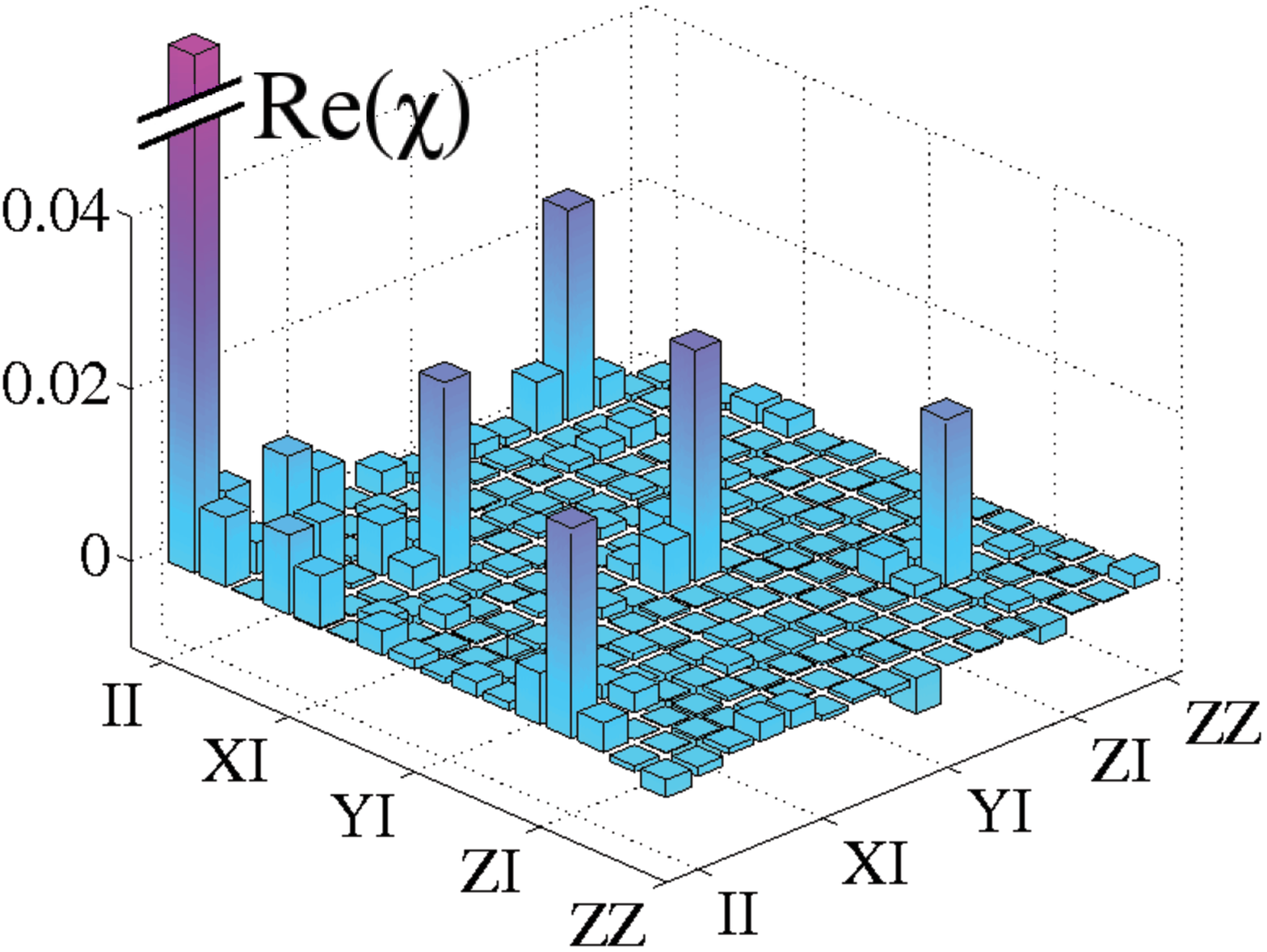}
  \includegraphics[width=4.15cm,angle=-00]{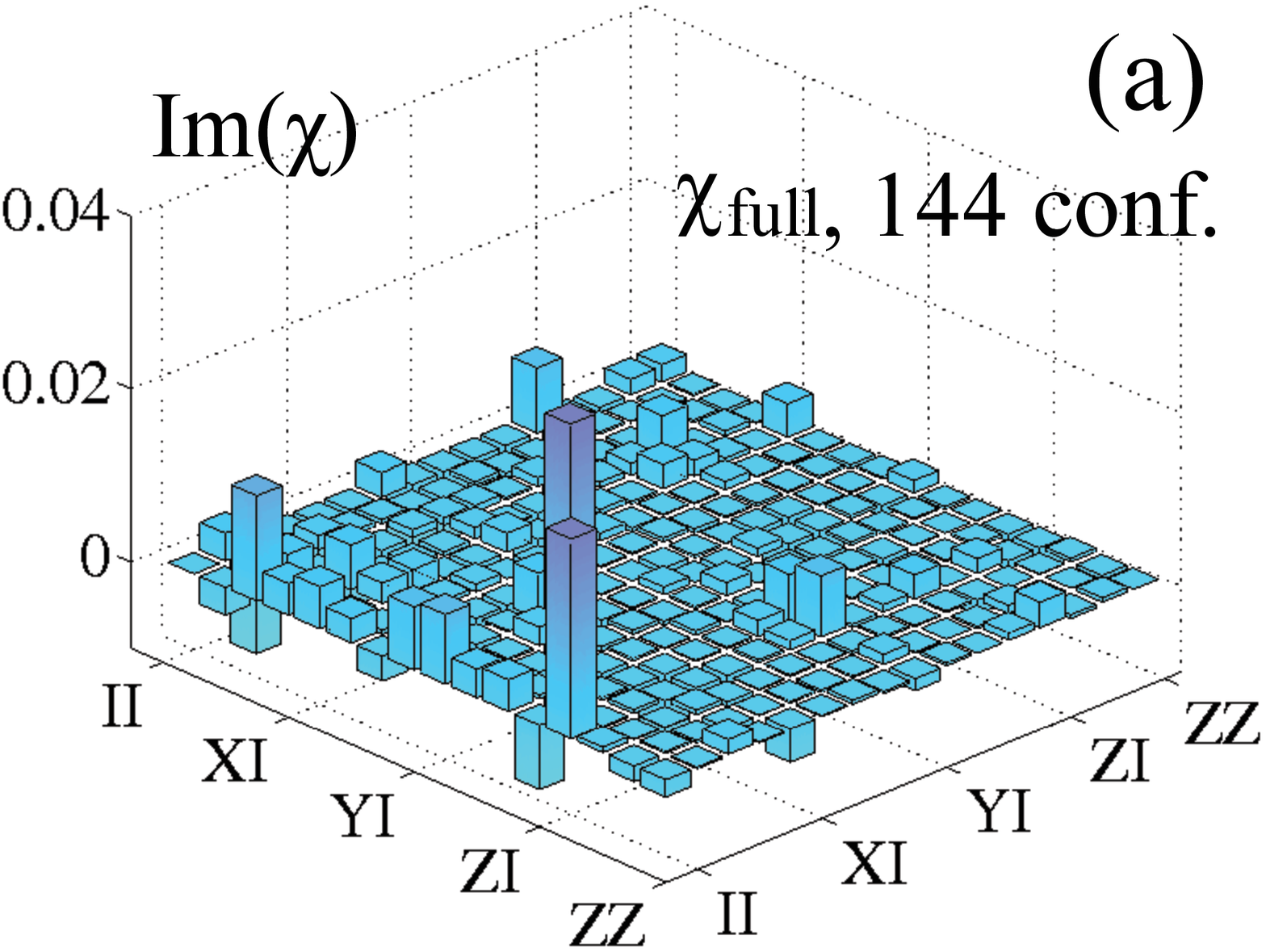}
  \includegraphics[width=4.15cm,angle=-00]{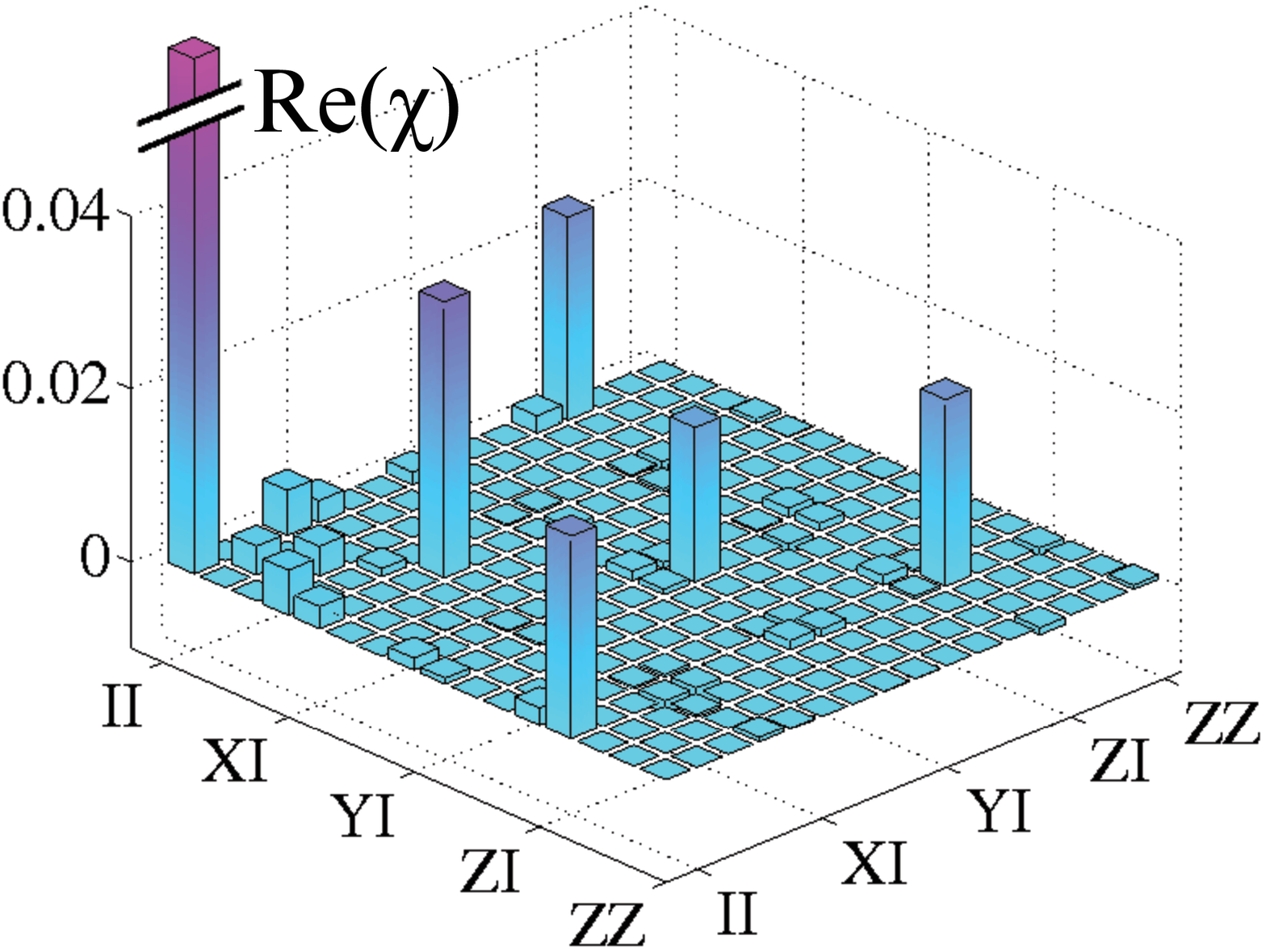}
  \includegraphics[width=4.15cm,angle=-00]{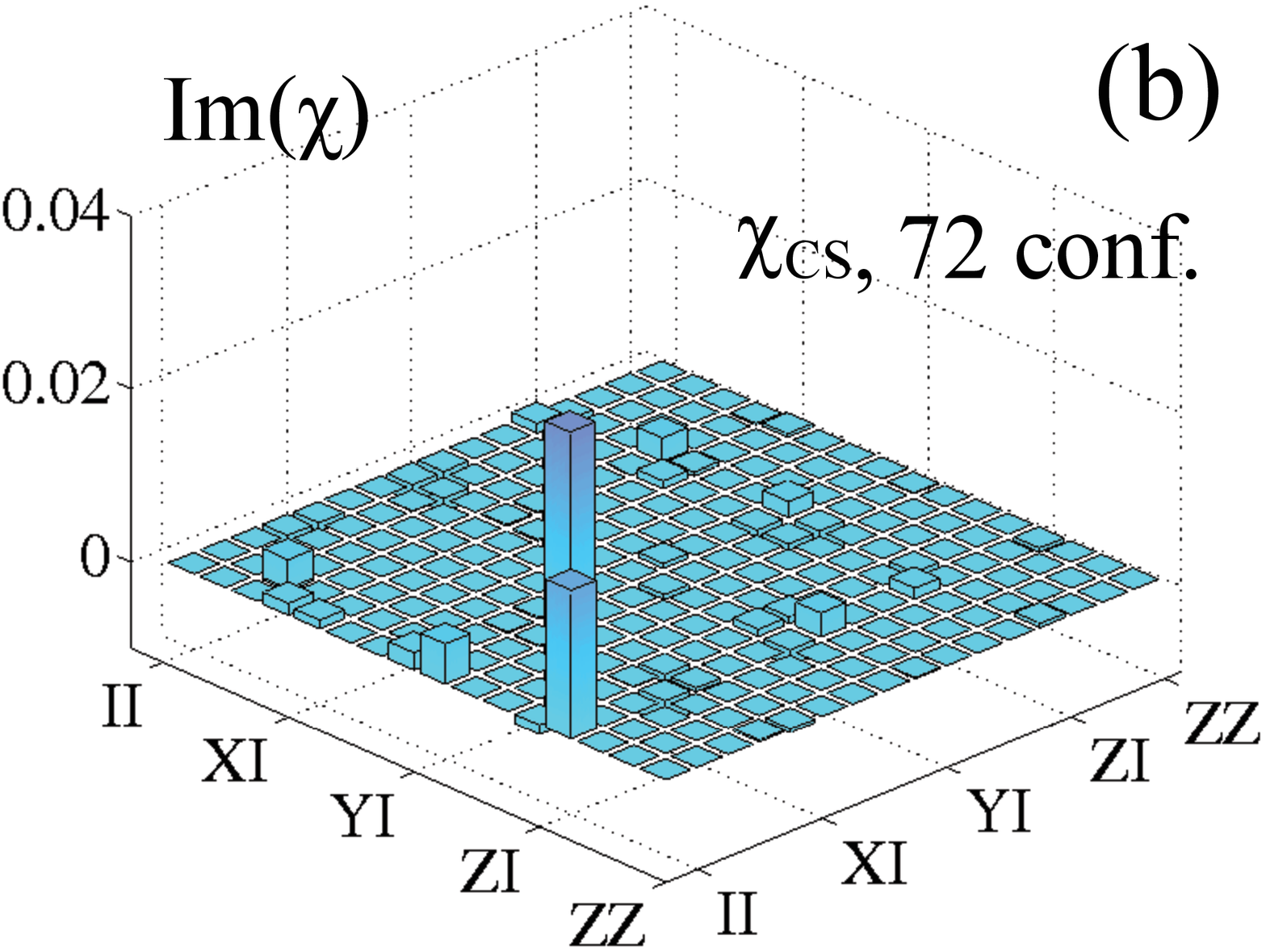}
  \includegraphics[width=4.15cm,angle=-00]{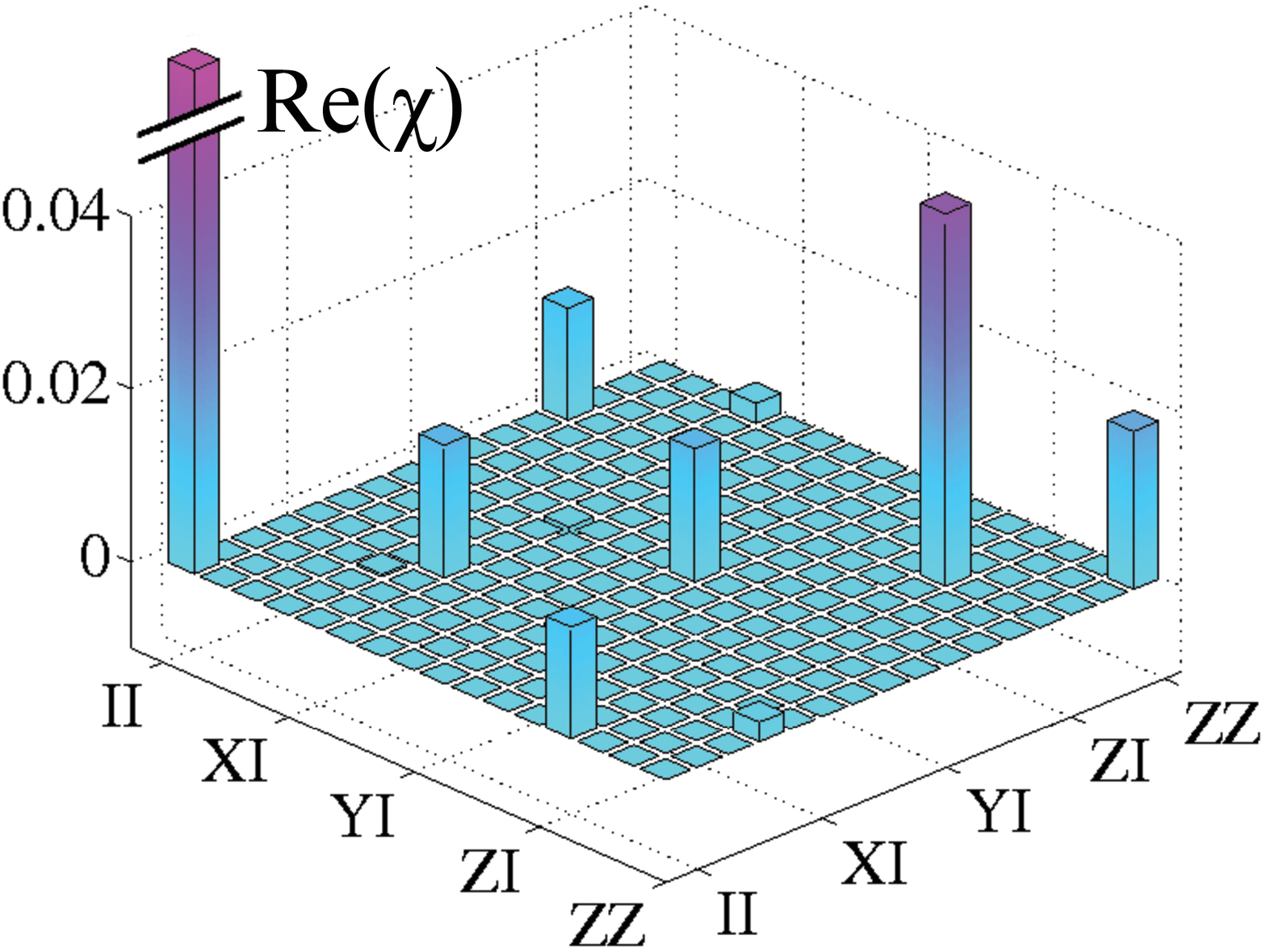}
  \includegraphics[width=4.15cm,angle=-00]{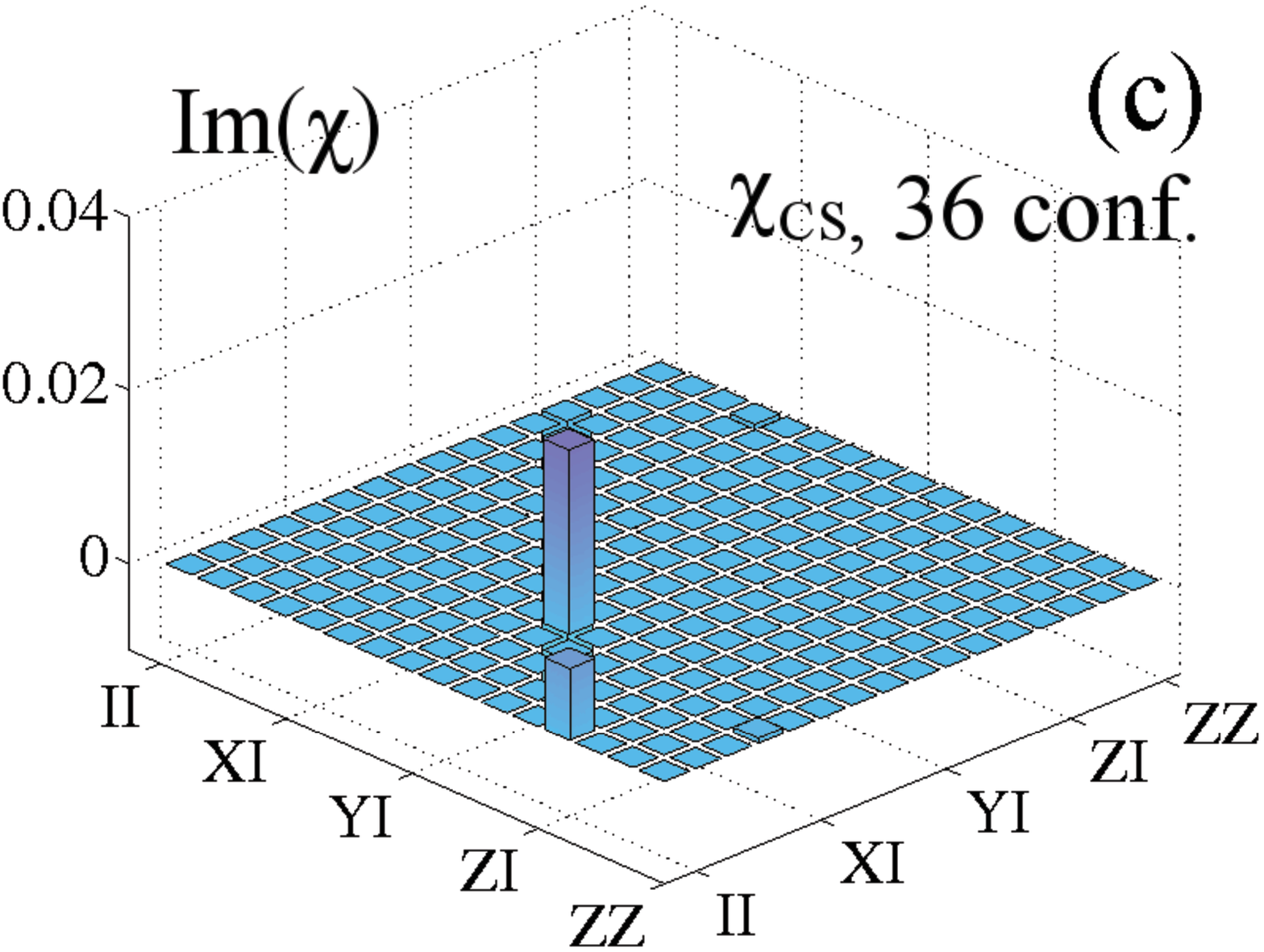}
\caption{(color online) (a) The process matrix $\chi_{\rm full}$
based on the full data set (144 configurations) and (b,c) the
CS-estimated matrices $\chi_{\rm CS}$ using a reduced data set: 72
configurations (b) and 36 configurations (c). The process matrices
are shown in the Pauli-error basis. The main element $\chi_{II,II}$
(process fidelity) is off the scale and therefore is cut; its height
is 0.907, 0.918, and 0.899 for the panels (a), (b), and (c),
respectively. All other peaks characterize imperfections. The
fidelity $F(\chi_{\rm CS}, \chi_{\rm full})$ for the matrices in
panels (b) and (c) is equal to 0.981 and 0.968, respectively.  The
middle and lower panels use the data set, corresponding to
underdetermined systems of equations.
  }
 \label{fig-chi}
\end{figure}

Figure \ref{fig-chi} shows examples of the CS estimated process
matrices $\chi_{CS}$ for $m_{\rm conf}=72$ (middle panel) and
$m_{\rm conf}=36$ (lower panel), together with the full-data process
matrix $\chi_{\rm full}$ (upper panel). The process matrices are
drawn in the Pauli-error basis to display the process imperfections
more clearly. The peak $\chi_{II,II}$ is off the scale and is cut
arbitrarily. We see that the CS estimated process matrices are
different from the full-data matrix; however the positions of the
main peaks are reproduced exactly, and their heights are also
reproduced rather well (for a small number of selected
configurations the peaks sometimes appear at wrong positions). It is
interesting to see that the CS procedure suppressed the height of
minor peaks. Note that both presented $\chi_{\rm CS}$ are based on
the data sets corresponding to underdetermined system of equations.

The computer resources needed for the calculation of results presented in Fig.\ 2 are not demanding. The calculations require about 30 MB  of computer memory and 2--4 seconds time for a modest PC per individual calculation (smaller time for smaller  number of configurations).

\begin{figure}[tb]
\includegraphics[width=8.5cm]{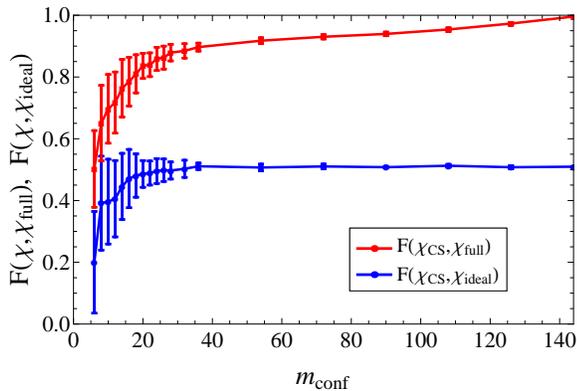}
\caption{(color online) Similar to Fig.\ \ref{fig2}, but for the CZ
gate realized with superconducting phase qubits. The process
fidelity $F(\chi_{\rm full},\chi_{\rm ideal})=0.51$  is much lower
than that for the Xmon qubit gate. As we see, CS QPT works
significantly better for this lower-fidelity gate than for the
better gate presented in Fig.\ \ref{fig2}.}
 \label{fig-phase}
\end{figure}

Besides the presented results, we have also performed analysis for
the CS QPT of two CZ gates based on phase qubits. The results are
qualitatively similar, except the process fidelity for phase-qubit
gates was significantly lower: 0.62 and 0.51. The results for one of
these gates are presented in Fig.\ \ref{fig-phase}. Comparing with
Fig.\ \ref{fig2}, we see that CS QPT works better for this
lower-fidelity gate. In particular, the blue line in Fig.\
\ref{fig-phase} is practically flat down to $m_{\rm conf}\simeq 20$
and the error bars are quite small. We think that the CS QPT works
better for a lower-fidelity gate because experimental imperfections
affect the measurement error relatively less in this case than for a
higher-fidelity gate.

Thus, our results show that for a CZ gate realized with
superconducting qubits CS QPT can reduce the number of used QPT
configurations by up to a factor of 7 compared with full QPT, and up
to a factor of 4 compared with the threshold at which the system of
equations for the standard QPT becomes underdetemined.

\subsection{Reduced data set, nonoptimal $\varepsilon$ }

\begin{figure}[tb]
\centering
\includegraphics[width=8.5cm]{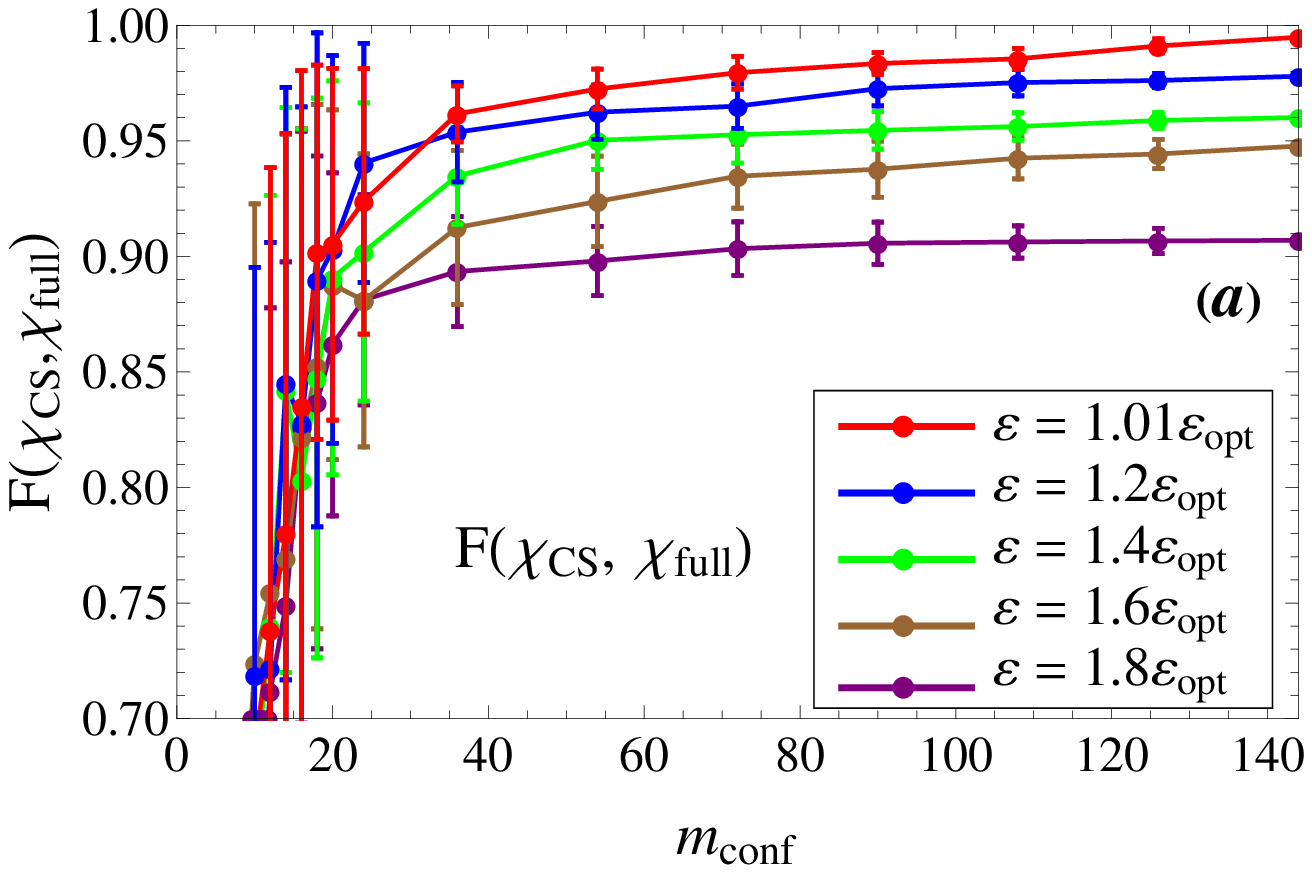}
\includegraphics[width=8.5cm]{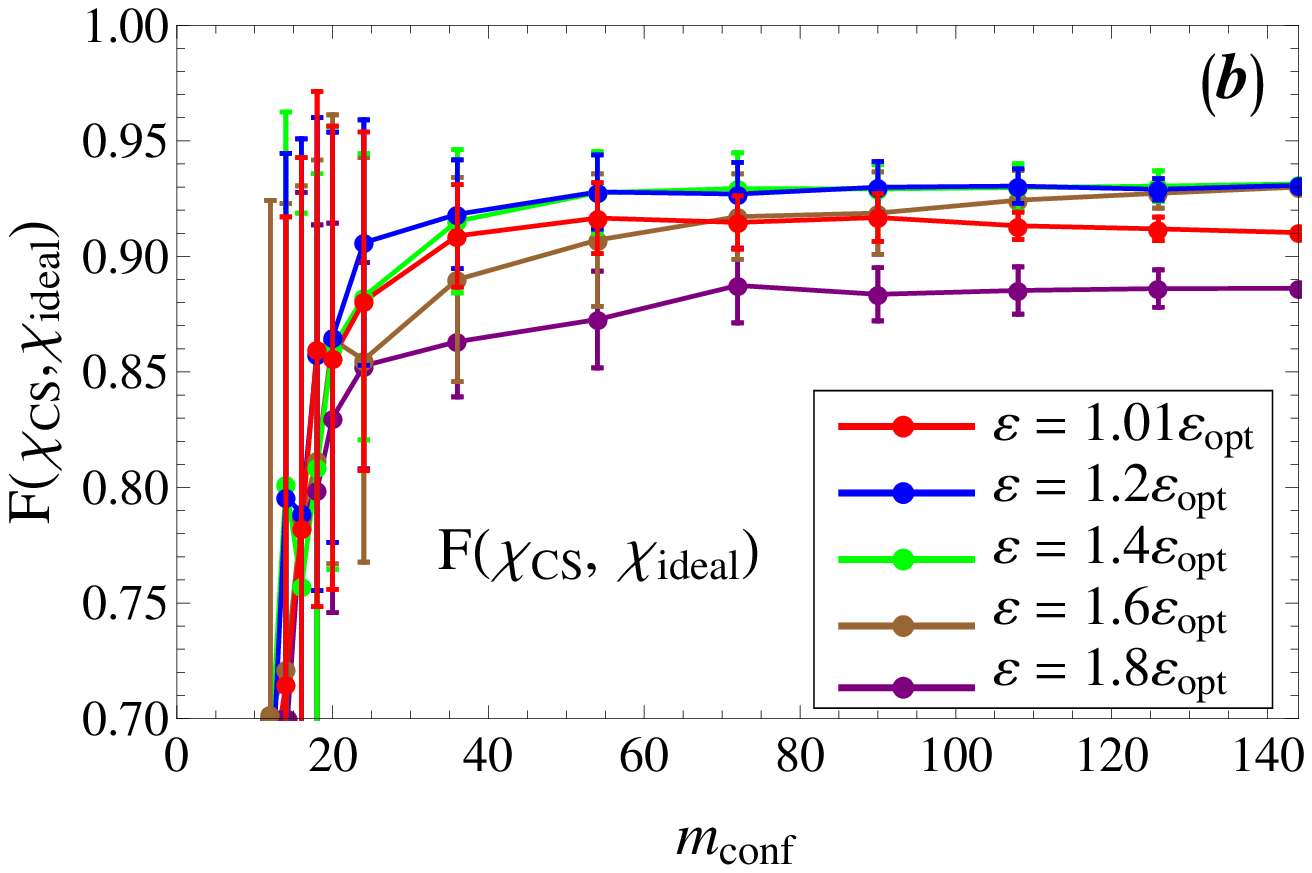}
\caption{(color online) (a) Fidelity $F(\chi_{\rm CS},\chi_{\rm
full})$ of the process matrix estimation and (b) the estimated
process fidelity $F(\chi_{\rm CS},\chi_{\rm ideal})$ as functions of
the data set size for several values of the noise parameter
$\varepsilon$ used in the CS optimization:
$\varepsilon/\varepsilon_{\rm opt}=1.01$, 1.2, 1.4, 1.6, and 1.8.
Error bars show the standard deviations calculated using 50 random
selections of reduced data sets. The red lines are the same as the
lines in Fig.\ \ref{fig2}.
   }
 \label{fig-nonopt}
\end{figure}

As mentioned above, in a QPT experiment with a reduced data set,
there is no straightforward way to find the near-optimal value of
the noise parameter $\varepsilon$ (which we find here from the full
data set). Therefore, it is important to check how well the CS
method works when a nonoptimal value of $\varepsilon$ is used.
Figure \ref{fig-nonopt} shows the results similar to those in Fig.\
\ref{fig2}, but with several values of the noise parameter:
$\varepsilon/\varepsilon_{\rm opt}=1.01$, 1.2, 1.4, 1.6, and 1.8.
The upper panel shows the fidelity between the matrix $\chi_{\rm
CS}$ and the full-data matrix $\chi_{\rm full}$; the lower panel
shows the process fidelity $F(\chi_{\rm CS}, \chi_{\rm ideal})$. We
see that the fidelity of the $\chi$ matrix estimation, $F(\chi_{\rm
CS}, \chi_{\rm full})$, becomes monotonously worse with increasing
$\varepsilon$, while the estimated process fidelity, $F(\chi_{\rm
CS}, \chi_{\rm ideal})$, may become larger when a nonoptimal
$\varepsilon$ is used.

Similar results (not presented here) for the CZ gate based on phase
qubits (see Fig.\ \ref{fig-phase}) have shown significantly better
tolerance to a nonoptimal choice  of $\varepsilon$; in particular,
even for $\varepsilon =3 \varepsilon_{\rm opt}$ the process fidelity
practically coincides with the blue line in Fig.\ \ref{fig-phase}
(obtained for $\varepsilon \approx \varepsilon_{\rm opt}$). We
believe the lower gate fidelity for phase qubits is responsible for
this relative insensitivity to the choice of $\varepsilon$.

\subsection{Comparison between Pauli-error and SVD bases}

So far for the CS method we have used the Pauli-error basis, in
which the process matrix $\chi$ is expected to be approximately
sparse because the ideal process matrix $\chi_{\rm ideal}$ contains
only one non-zero element, $\chi_{{\rm ideal}, II,II}=1$. However,
there are an infinite number of the operator bases with this
property: for example, the SVD basis (see Appendix B) suggested in
Refs.\ \cite{KosutSVD} and \cite{ShabaniKosut}. The process matrix
is different in the Pauli-error and SVD bases, therefore the CS
optimization should produce different results. To compare the
results, we do the CS optimization in the SVD basis, then convert
the resulting matrix $\chi$ into the Pauli-error basis, and
calculate the fidelity $F(\chi_{\rm CS-SVD},\chi_{\rm CS})$ between
the transformed process matrix and the matrix $\chi_{\rm CS}$
obtained using optimization in the Pauli-error basis directly.

\begin{figure}[tb]
\includegraphics[width=8.5cm]{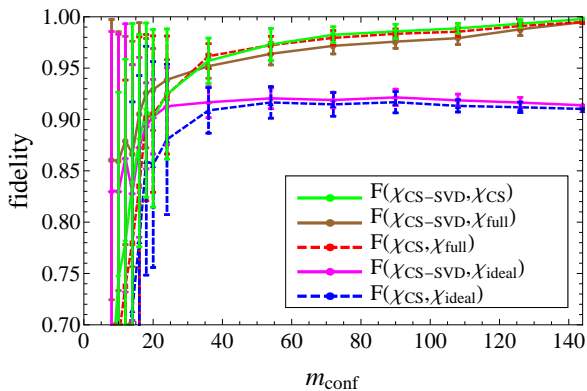}
\caption{(color online) Comparison between the CS results obtained
in the SVD and Pauli-error bases. The green line shows the relative
fidelity $F(\chi_{\rm CS-SVD}, \chi_{\rm CS})$ as a function of the
number $m_{\rm conf}$ of randomly selected configurations. We also
show the fidelities $F(\chi_{\rm CS-SVD}, \chi_{\rm full})$ (brown
line), $F(\chi_{\rm CS}, \chi_{\rm full})$ (red dashed line), and
process fidelities  $F(\chi_{\rm CS-SVD},\chi_{\rm ideal})$ (magenta
line) and $F(\chi_{\rm CS},\chi_{\rm ideal})$ (blue dashed line).
The dashed lines have been shown in Fig.\ \ref{fig2}. The results
using the SVD basis are somewhat more accurate than those for the
Pauli-error basis when $m_{\rm conf} <40$.
 }
 \label{comp-SVD-Pauli}
\end{figure}

The green line in Fig.\ \ref{comp-SVD-Pauli} shows $F(\chi_{\rm
CS-SVD},\chi_{\rm CS})$ as a function of the selected size of the
data set for the CZ gate realized with Xmon qubits, similar to Fig.\
\ref{fig2} (the same $\varepsilon$ is used). We also show the
fidelity between the SVD-basis-obtained matrix $\chi_{\rm CS-SVD}$
and the full-data matrix $\chi_{\rm full}$ as well as the ideal
process matrix  $\chi_{\rm ideal}$. For comparison we also include
the lines shown in Fig.\ \ref{fig2} (dashed lines), obtained using the Pauli-error
basis. As we see, the results obtained in the two bases are close to
each other, though the SVD basis seems to work a little better at
small data sizes, $m_{\rm conf}\simeq 20$. The visual comparison of
$\chi$-matrices obtained in these bases (as in Fig.\ \ref{fig-chi},
not presented here) also shows that they are quite similar. It
should be noted that the calculations in the SVD basis are somewhat
faster ($\sim$2 seconds per point) and require less memory ($\sim$6
MB) than the calculations in the Pauli-error basis. This is because
the matrix $\Phi$ defined in Eq.\ (\ref{Vectorized}) for the CZ gate
contains about half the number of non-zero elements in the SVD basis
than in the Pauli-error basis.

All results presented here are obtained using the CVX-SeDuMi
package. The results for the CZ gate obtained using the YALMIP-SDPT3
package are similar when the same value of $\varepsilon$ is used.
Surprisingly, in our realization of computation, the YALMIP-SDPT3
package still finds reasonable solutions when $\varepsilon$ is
significantly smaller than $\varepsilon_{\rm opt}$ (even when
$\varepsilon$ is zero or negative), so that the problem cannot have
a solution; apparently in this case the solver increases the value
of $\varepsilon$ until a solution is found. This may seem to be a
good feature of YALMIP-SDPT3. However, using $\varepsilon <
\varepsilon_{\rm opt}$ should decrease the accuracy of the result
(see the next subsection). Moreover, YALMIP-SDPT3 does not work well
for the Toffoli gate discussed in Sec.\ \ref{OurResults-ThreeQu}.
Thus we conclude that CVX-SeDuMi package is better than YALMIP-SDPT3
package for our CS calculations. (Note that this finding may be
specific to our system.)

\subsection{Comparison with least-squares minimization}
\label{Sec-2q-LS}

\begin{figure}[tb]
\includegraphics[width=8.5cm]{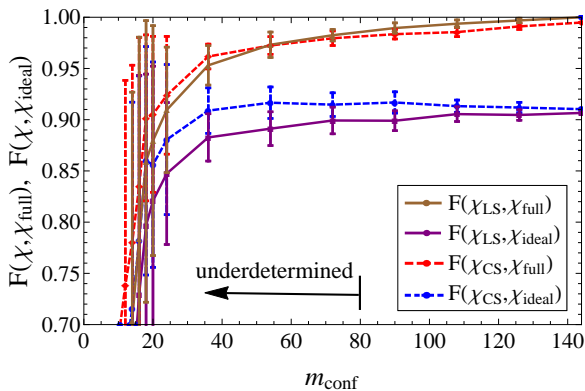}
\caption{(color online) Comparison between the results obtained by
the LS and CS methods. The solid lines are for the LS method, the
dashed lines (same as in Fig.\ \ref{fig2}) are for the CS method.
The CS method is more accurate for a substantially reduced data set.
   }
 \label{fig-LS-2q}
\end{figure}

Besides using the CS method for reduced data sets, we also used the
LS minimization [with constraints (\ref{ChiPositive}) and
(\ref{ChiTracePreserving})] for the same reduced sets. Solid lines
in Fig.\ \ref{fig-LS-2q} show the resulting fidelity $F(\chi_{\rm
LS},\chi_{\rm full})$ compared with the full-data process matrix and
the estimated process fidelity $F(\chi_{\rm LS},\chi_{\rm ideal})$.

Somewhat surprisingly, the LS method still works (though less well)
in a significantly underdetermined regime. Naively, we would expect
that in this case Eq.\ (\ref{Vectorized}) can be satisfied exactly,
and there are many exact solutions corresponding to the null space
of the selected part of the matrix $\Phi$. However, numerical
results show that in reality Eq.\ (\ref{Vectorized}) cannot be
satisfied exactly unless the selected data set is very small. The
reason is that the matrix $\chi$ has to be positive, and the
(corrected) experimental probabilities can be close to the limits of
the physical range or even outside it.

The problem is that the experimental probabilities are not directly
obtained from the experiment, but are corrected for imperfect
measurement fidelity \cite{Mariantoni-11}. As a result, they may
become larger than one or smaller than zero. This happens fairly
often for high fidelity gates because for an ideal operation the
measurement results are often zeros and ones, so the experimental
probabilities should also be close to zero or one. Any additional
deviation due to imperfect correction for the measurement fidelity
may then push the probabilities outside of the physical range. It is obvious
that in this case Eq.\ (\ref{Vectorized}) cannot be satisfied
exactly for any physical $\chi$. To resolve this problem one could
consider rescaling the probabilities in such instances, so that they
are exactly one or zero instead of lying outside the range. However,
this also does not help much because a probability of one means that
the resulting state is pure, so this strongly reduces the number of free
parameters in the process matrix $\chi$. As a result, Eq.\
(\ref{Vectorized}) cannot be satisfied exactly, and the LS
minimization is formally possible even in the underdetermined case.

Another reason why Eq.\ (\ref{Vectorized}) may be impossible to
satisfy in the underdetermined case, is that the randomly selected
rows of the matrix $\Phi$ can be linearly dependent. Then
mathematically some linear relations between the experimental
probabilities must be satisfied, while in reality they are obviously
not satisfied exactly.

These reasons make the LS minimization a mathematically possible
procedure even in the underdetermined regime. However, as we see
from Fig.\ \ref{fig-LS-2q}, in this case the procedure works less
well than the compressed sensing, estimating the process matrix and
process fidelity with a lower accuracy. Similar calculations for the
CZ gate realized with phase qubits (not presented here) also show
that the LS method does not work well at relatively small $m_{\rm
conf}$.
 The advantage of the
compressed sensing in comparison with the LS minimization becomes
even stronger for the three-qubit Toffoli gate considered in the
next section. Note though that when the selected data set is large
enough to give an overdetermined system of equations
(\ref{Vectorized}), the LS method works better than the CS method.
Therefore, the compressed sensing is beneficial only for a
substantially reduced (underdetermined) data set, which is exactly
the desired regime of operation.

\section{Three-Qubit CS QPT for Toffoli gate}
\label{OurResults-ThreeQu}

In this section we apply the compressed sensing method to simulated
tomographic data corresponding to a three-qubit Toffoli gate
\cite{N-C,CoryToffoli,MonzRealizeToffoli,Mariantoni-11,FedorovImplementToffoli}.
As discussed in Sec.\ \ref{Configurations}, the process matrix of a
three-qubit gate contains $16^3-4^3 = 4032$ independent real
parameters, while the full QPT requires  $M_{\rm conf} =12^3= 1728$
measurement configurations yielding a total of $M = 12^3\times 2^3 =
13824$ experimental probabilities, if we use $n_{\rm in}=4$ initial
states and $n_{\rm R}=3$ measurement rotations per qubit, with all
qubits measured independently. If we work with a partial data set,
the system of equations (\ref{Vectorized}) becomes underdetermined
if the number $m_{\rm conf}$ of used configurations is less than
$4032/7=576$. In such a regime the traditional maximum likelihood or
LS methods are not expected to provide a good estimate of the process
matrix. In this section we demonstrate that for our simulated
Toffoli gate the compressed sensing method works well even for a
much smaller number of configurations, $m_{\rm conf}\ll 576$.

For the analysis we have simulated experimental data corresponding
to a noisy Toffoli gate by adding truncated Gaussian noise with
a small amplitude to each of $M=13824$ ideal measurement
probabilities~$P^{\rm {ideal}}_i$.  We assumed the set of
experimental probabilities in Eq.\ (\ref{Vectorized}) to be of the
form $P^{\textrm{exp}}_i = P^{\textrm{ideal}}_i + \Delta{P}_i$,
where $\Delta{P}_{i}$ are random numbers sampled from the normal
distribution with zero mean and a small standard deviation $\sigma$.
By choosing different values of the standard
deviation $\sigma$ we can change the process fidelity of the simulated
Toffoli gate: a smaller value of $\sigma$ makes the process fidelity
closer to 1. After adding the Gaussian
noise~$\Delta{P}_i$ to the ideal probabilities~$P^{\rm ideal}_i$, we
check whether the resulting simulated
probabilities~$P^{\textrm{exp}}_i$ are in the interval $ [0,1]$. If
a~$P^\textrm{{exp}}_i$ happens to be outside the interval $[0,1]$,
we repeat the procedure until the condition $P^{\textrm{exp}}_i\in
[0,1]$ is satisfied. Finally, we renormalize each  set of 8
probabilities corresponding to the same measurement configuration so
that these probabilities add up to~$1$.

Thus the simulated imperfect quantum process is defined by $M=13824$
probabilities, corresponding to $M_{\rm conf}=1728$ configurations;
the process fidelity for a particular realization (used here) with
$\sigma=0.01$ is $F_\chi=F(\chi_{\rm full},\chi_{\rm
ideal})=0.959$. We then test efficiency of the compressed sensing
method by randomly selecting $m_{\rm conf}\leq 1728$ configurations,
finding the corresponding process matrix $\chi_{\rm CS}$, and
comparing it with the full-data matrix $\chi_{\rm full}$ by
calculating the fidelity $F(\chi_{\rm CS},\chi_{\rm full})$. We also
calculate the process fidelity $F(\chi_{\rm CS},\chi_{\rm ideal})$
given by $\chi_{\rm CS}$.

\begin{figure}[tb]
\includegraphics[width=8.5cm]{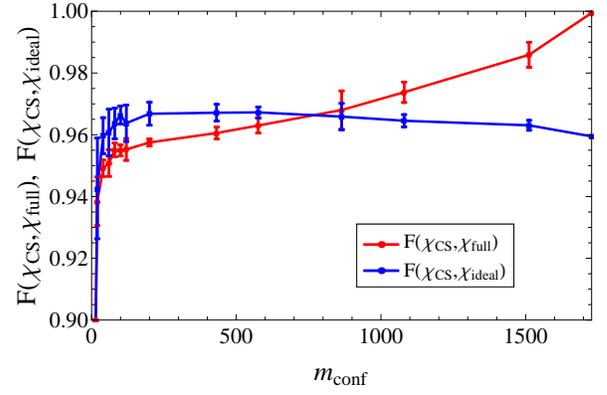}
\caption{(color online) CS QPT for a simulated Toffoli gate. Red
line: fidelity $F(\chi_{\rm CS}, \chi_{\rm full})$ of the process
matrix estimation, blue line: the estimated process fidelity
$F(\chi_{\rm CS}, \chi_{\rm ideal})$, both as functions of the data
set size, expressed as the number $m_{\rm conf}$ of randomly selected
configurations. The full QPT corresponds to 1728 configurations. The
system of equations becomes underdetermined when $m_{\rm conf} <
576$.
  }
 \label{Fig-3q-CS}
\end{figure}

The red line in Fig.\ \ref{Fig-3q-CS} shows the fidelity
$F(\chi_{\rm CS},\chi_{\rm full})$ as a function of the number
$m_{\rm conf}$ of randomly selected configurations. The value of
$\varepsilon$ is chosen to be practically equal to $\varepsilon_{\rm
opt}=||(\vec{P}^{\rm exp}_{\rm full} - \Phi\vec{\chi}_{\rm
full})||_{\ell_2}/\sqrt{M}=0.01146$ (the relative difference is less
than $10^{-3}$). The $\ell_1$-minimization is done using the
CVX-SeDuMi package. The error bars are calculated by repeating the
procedure of random selection 7 times. We see a reasonably high
fidelity $F(\chi_{\rm CS},\chi_{\rm full})$ of the reconstructed
process matrix even for small numbers of selected configurations.
For example, $F(\chi_{\rm CS},\chi_{\rm full})=0.95$ for only
$m_{\rm conf}=40$ configurations, which represents a reduction by
more than a factor of 40 compared with the full QPT and
approximately a factor of 15 compared with the threshold of the
underdetermined system of equations.

The blue line in Fig.\ \ref{Fig-3q-CS} shows the process fidelity
$F(\chi_{\rm CS},\chi_{\rm ideal})$ calculated by the CS method. We
see that it remains practically flat down to $m_{\rm conf}\agt 40$,
which means that $\chi_{\rm CS}$ can be used efficiently to estimate
the actual process fidelity.

\begin{figure}[tb]
\includegraphics[width=8.5cm]{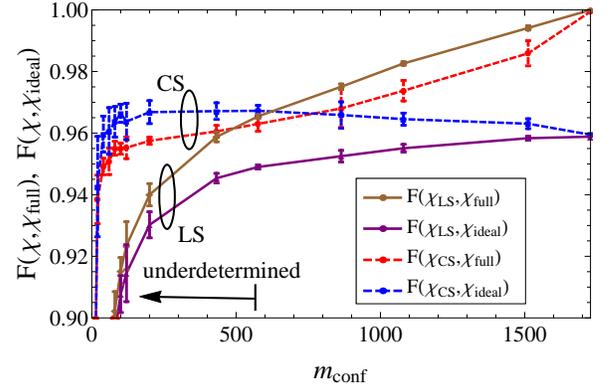}
\caption{(color online) Comparison between the calculations using CS
and LS methods for the simulated Toffoli gate. Solid lines are for
the LS method, dashed lines (the same as in Fig.\ \ref{Fig-3q-CS})
are for the CS method. In the underdetermined regime the CS-method
results are much better than the LS-method results.
  }
 \label{Fig9}
\end{figure}

Figure \ref{Fig9} shows similar results calculated using the LS
method (for comparison the lines from Fig.\ \ref{Fig-3q-CS} are
shown by dashed lines). We see that the LS method still works in the
underdetermined regime ($m_{\rm conf}<576$); however, it works
significantly worse than the CS method. As an example, for $m_{\rm
conf}=40$ the fidelity of the process matrix estimation using the LS
method is $F(\chi_{\rm LS},\chi_{\rm full})=0.86$, which is
significantly less than $F(\chi_{\rm CS},\chi_{\rm full})=0.95$ for
the CS method. Similarly, for $m_{\rm conf}=40$ the process fidelity
obtained via the CS method, $F(\chi_{\rm CS},\chi_{\rm ideal})=0.96$
is close to the full-data value of 0.959, while the LS-method value,
$F(\chi_{\rm LS},\chi_{\rm ideal})=0.85$, is quite different.

Besides using the Pauli-error basis for the results shown in Fig.\ \ref{Fig-3q-CS}, we have also performed the calculations using the SVD basis. The results (not shown) are very close to those in Fig.\ \ref{Fig-3q-CS}, and the relative fidelity $F(\chi_{\rm CS-SVD}, \chi_{\rm CS})$ is above 0.98 for $m_{\rm conf}>200$ and above 0.95 for $m_{\rm conf}>40$. We have also performed the calculations using non-optimal values of the noise parameter $\varepsilon$. In comparison with the results for CZ gate shown in Fig.\ \ref{fig-nonopt}, the results for the Toffoli gate (not shown) are more sensitive to the variation of $\varepsilon$. In particular, the fidelity $F(\chi_{\rm CS}, \chi_{\rm full})$  is about 0.93 for $\varepsilon =1.2 \varepsilon_{\rm opt}$ (not significantly depending on $m_{\rm conf}$ for $m_{\rm conf}>40$) and the process fidelity $F(\chi_{\rm CS}, \chi_{\rm ideal})$ for $\varepsilon =1.2 \varepsilon_{\rm opt}$ is approximately 0.93 instead of the actual value 0.96.

Compared with the two-qubit case, it takes significantly more
computing time and memory to solve the $\ell_{1}$-minimization
problem for three qubits. In particular, our calculations in the
Pauli-error basis took about 8 hours per point on a personal
computer for $m_{\rm conf}\simeq 1500$ and about 1.5 hours per point
for $m_{\rm conf}\simeq 40$; this is three orders of magnitude
longer than for two qubits. The amount of used computer memory was
3--10 GB, which is two orders of magnitude larger than for two
qubits. (The calculations in the SVD basis for the Toffoli gate took
1--3 hours per point and $\sim$2 GB of memory.) Such a strong
scaling of required computer resources with the number of qubits
seems to be the limiting factor in extending the CS QPT beyond three
qubits, unless a more efficient algorithm is found. (Note that LS calculations required similar amount of memory, but the computation time was much shorter.)

The presented results have been obtained using the CVX-SeDuMi
package. We also attempted to use the YALMIP-SDPT3 package. However, in our realization of computation
the calculation results were very unreliable  for $m_{\rm conf}<200$
using the SVD basis, and even worse when the Pauli-error basis was
used. Therefore we decided to use only the CVX-SeDuMi package for
the 3-qubit CS procedure.

\section{Standard deviation of state fidelity}
\label{FidSquaredSection}

As shown in previous sections, the process matrices $\chi_{\rm CS}$ obtained via the CS method allow us to estimate reliably the process fidelity $F_\chi = F(\chi, \chi_{\rm ideal})$ of a gate using just a small fraction of the full experimental data. While $F_\chi$ is the most widely used characteristic of an experimental gate accuracy, it is not the only one. An equivalent characteristic (usually used in randomized benchmarking) is the average state fidelity, defined as $\overline{F_{\rm st}}=\int {\rm Tr} (\rho_{\rm actual} \rho_{\rm ideal})\, d  |\psi_{\rm in}\rangle / \int d |\psi_{\rm in}\rangle$, where the integration is over the initial pure states $|\psi_{\rm in}\rangle$ (using the Haar measure; it is often assumed that $\int d |\psi_{\rm in}\rangle=1$), while the states $\rho_{\rm ideal}$ and $\rho_{\rm actual}$ are the ideal and actual final states for the initial state $|\psi_{\rm in}\rangle$. The average state fidelity $\overline{F_{\rm st}}$ is
sometimes called the ``gate fidelity'' \cite{Magesan-12} and can be naturally measured in the randomized benchmarking ($F_{\rm RB}=\overline{F_{\rm st}}$); it is linearly related ~\cite{Horodecki,NielsenAveGateFidelity} to the process fidelity, $\overline{F_{\rm st}}=(F_\chi d +1)/(d+1)$, where $d=2^N$ is the Hilbert space dimension.

Besides the average state fidelity, an obviously important characteristic of a gate operation is the worst-case state fidelity $F_{\rm st, min}$, which is minimized over the initial state. Unfortunately, the minimum state fidelity is hard to find computationally even when the process matrix $\chi$ is known. Another natural characteristic is the standard deviation of the state fidelity,
    \be
    \Delta F_{\rm st}=\sqrt{\overline{F^2_{\rm st}}-\overline{F_{\rm st}}^{\,2}},
    \label{std-def}\ee
where $\overline{F_{\rm st}^2}=\int [{\rm Tr} (\rho_{\rm actual}
\rho_{\rm ideal})]^2\, d  |\psi_{\rm in}\rangle / \int d |\psi_{\rm
in}\rangle$ is the average square of the state fidelity. The
advantage of $\Delta F_{\rm st}$ in comparison with $F_{\rm st,
min}$ is that $\overline{F_{\rm st}^2}$ and $\Delta F_{\rm st}$ can
be calculated from $\chi$ in a straightforward way \cite{Molmer2008,
Emerson2011}. Our way of calculating  $\overline{F_{\rm st}^2}$ is
described in Appendix C [see Eq.\ (\ref{fidsquared})].

\begin{figure}[tb]
\includegraphics[width=8.5cm]{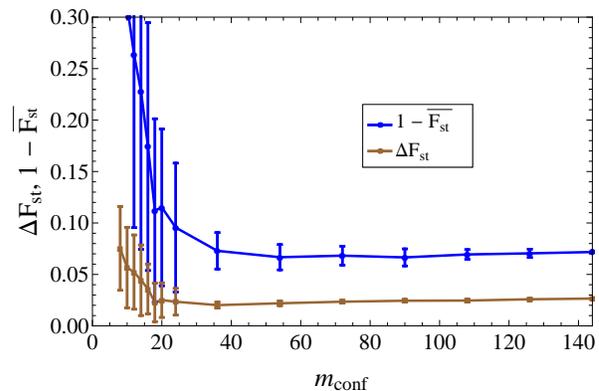}
  \caption{(color online) Blue (upper) line: average state infidelity $1-\overline{F_{\rm st}}$ for the CS-estimated process matrix $\chi_{\rm CS}$ as a function of the selected data set size for the experimental CZ gate (this line is linearly related to the blue line in Fig.\ \ref{fig2}). Brown (lower) line: the standard deviation of the state fidelity $\Delta F_{\rm st}$, defined via variation of the initial state, Eq.\ (\ref{std-def}), using the same $\chi_{\rm CS}$. The error bars are computed by repeating the procedure 50 times with different random selections of used configurations.  }
 \label{Fig-std-2q}
\end{figure}

\begin{figure}[tb]
\includegraphics[width=8.5cm]{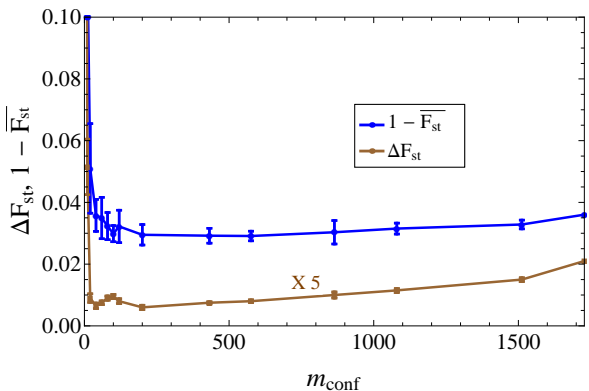}
  \caption{(color online) The same as in Fig.\ \ref{Fig-std-2q}, but for the simulated Toffoli gate. The random selection of configurations is repeated 7 times for each point. The results for the standard deviation $\Delta F_{\rm st}$ are multiplied by the
  factor of 5 for clarity.
  }
 \label{Fig-std-3q}
\end{figure}

We have analyzed numerically how well the CS QPT estimates $\Delta F_{\rm st}$ from the reduced data set, using the previously calculated process matrices $\chi_{\rm CS}$ for the experimental CZ gate and the simulated Toffoli gate (considered in Secs.\ \ref{results-two-qubits} and \ref{OurResults-ThreeQu}). The results are presented in Figs.\ \ref{Fig-std-2q} and \ref{Fig-std-3q}. We show the average state infidelity, $1-\overline{F_{\rm st}}$, and the standard deviation of the state fidelity, $\Delta F_{\rm st}$, as functions of the number of selected configurations, $m_{\rm conf}$. The random selection of used configurations is repeated 50 times for Fig.\ \ref{Fig-std-2q} (7 times for Fig.\ \ref{Fig-std-3q}), the error bars show the statistical variation, while the dots show the average values.

As seen from Figs.\ \ref{Fig-std-2q} and \ref{Fig-std-3q}, the CS method estimates reasonably well not only the average state fidelity $\overline{F_{\rm st}}$ (which is equivalent to $F_{\chi}$ presented in Figs.\ \ref{fig2} and \ref{Fig-3q-CS}), but also its standard deviation  $\Delta F_{\rm st}$.
It is interesting to note that $\Delta F_{\rm st}$ is significantly smaller than the infidelity $1-\overline{F_{\rm st}}$, which means that the state fidelity ${\rm Tr} (\rho_{\rm acual}\rho_{\rm ideal})$ does not vary significantly for different initial states [the ratio $\Delta F_{\rm st}/(1-\overline{F_{\rm st}})$ is especially small for the simulated Toffoli gate, though this may be because of our particular way of simulation].

\section{Conclusion}\label{Conclusion}

In this paper we have numerically analyzed the efficiency of
compressed sensing quantum process tomography (CS QPT)
\cite{KosutSVD,ShabaniKosut} applied to superconducting qubits (we
did not consider the CS method of Refs.\
\cite{GrossFlammia2010,Flammia2012}). We have used experimental data
for two-qubit CZ gates realized with Xmon and phase qubits, and
simulated data for the three-qubit Toffoli gate with numerically
added noise. We have shown that CS QPT permits a reasonably high
fidelity estimation of the process matrix from a substantially
reduced data set compared to the full QPT. In particular, for the CZ
gate (Fig.\ \ref{fig2}) the amount of data can be reduced by a
factor of $\sim$7 compared to the full QPT (which is a factor of $\sim$4
compared to the threshold of underdetermined system of equations).
For the Toffoli gate (Fig.\ \ref{Fig-3q-CS}) the data reduction
factor is $\sim$40 compared to the full QPT ($\sim$15 compared to
the threshold of underdeterminacy).

In our analysis we have primarily used two characteristics. The
first characteristic is the comparison between the CS-obtained
process matrix $\chi_{\rm CS}$ and the matrix $\chi_{\rm full}$
obtained from the full data set; this comparison is quantitatively
represented by the fidelity $F(\chi_{\rm CS},\chi_{\rm full})$. The
second characteristic is how well the CS method estimates the
process fidelity $F_\chi$, i.e., how close $F(\chi_{\rm
CS},\chi_{\rm ideal})$ is to the full-data value $F(\chi_{\rm
full},\chi_{\rm ideal})$. Besides these two characteristics, we have
also calculated the standard deviation of the state fidelity $\Delta
F_{\rm st}$ [Eq.\ (\ref{std-def})] and checked how well the CS
method estimates $\Delta F_{\rm st}$ from a reduced data set (Figs.\
\ref{Fig-std-2q} and \ref{Fig-std-3q}). Our compressed sensing
method depends on the choice of the basis, in which the process
matrix should be approximately sparse, and also depends on the choice of
the noise parameter $\varepsilon$ [see Eq.\
(\ref{ConditionsL1Problem})]. We have used two bases: the
Pauli-error basis and the SVD basis. The results obtained in both
bases are similar to each other, though the SVD basis required less
computational resources. The issue of choosing proper $\varepsilon$
is not trivial. In our calculations we have used a value slightly
larger than the noise level calculated from the full data set.
However, in an experiment with a reduced data set this way of
choosing $\varepsilon$ is not possible, so its value should be
chosen from an estimate of the inaccuracy of the experimental
probabilities. We have shown that the CS method tolerates some
inaccuracy of $\varepsilon$ (up to $\sim$60$\%$ for the results
shown in Fig.\ \ref{fig-nonopt}); however, finding a proper way of
choosing $\varepsilon$ is still an open issue.

We have also compared the performance of the CS method with the
least squares optimization. Somewhat surprisingly, the LS method can
still be applied when the systems of equations (\ref{Vectorized}) is
underdetermined (unless the data set size is too small). This is
because the condition of a process matrix being physical (positive,
trace-preserving) usually makes satisfying Eqs.\ (\ref{Vectorized})
impossible. However, even though the LS method formally works, it
gives a less accurate estimate of $\chi$ than the CS method in the
significantly underdetermined regime (although it does give a better
estimate in the overdetermined regime). The advantage of the CS
method over the LS method is more pronounced for the Toffoli gate
(Fig.\ \ref{Fig9}).

Thus the CS QPT is useful  for two-qubit and three-qubit quantum
gates based on superconducting qubits. The method offers a very
significant reduction of the needed amount of experimental data.
However, the scaling of the required computing resources with the
number of qubits seems to be prohibitive: in our calculations it
took three orders of magnitude longer and two orders of magnitude
more memory for the three-qubit-gate calculation than for two
qubits. Such a scaling of computing resources seems to be a limiting
factor in the application of our implementation of the CS method for
QPT of four or more qubits. Therefore, the development of more
efficient numerical algorithms for the CS QPT is an important task
for future research.

\section*{ACKNOWLEDGEMENTS}

The authors thank Yuri Bogdanov, Steven Flammia, Justin Dressel, and Eyob Sete for useful discussions. We also thank Matteo Mariantoni for being involved in this work at its early stage.
The research was funded by the
Office of the Director of National Intelligence (ODNI), Intelligence
Advanced Research Projects Activity (IARPA), through the Army
Research Office Grant No. W911NF-10-1-0334. All statements of fact,
opinion, or conclusions contained herein are those of the authors
and should not be construed as representing the official views or
policies of IARPA, the ODNI, or the U.S. Government. We also
acknowledge support from the ARO MURI Grant No. W911NF-11-1-0268.

\appendix

\section{Pauli-error basis}
\label{AppendixChiTilde}

In this Appendix we discuss the definition of the Pauli-error basis
used in this paper. The detailed theory of the QPT in the
Pauli-error basis is presented in Ref.\ \cite{Korotkov-13}.

Let us start with description of a quantum process $\mathcal{E}$ in
the Pauli basis $\{ {\cal P}_\alpha\}$,
    \be
\rho^{\rm in} \mapsto \mathcal{E}(\rho^{\rm in})=\sum_{\alpha,\beta
= 1}^{d^{2}}\chi_{\alpha \beta} {\cal P}_{\alpha}\rho^{\rm in} {\cal
P}_{\beta}^{\dagger},
    \label{AppA-Pauli}\ee
where for generality ${\cal P}$ is not necessarily Hermitian (to
include the modified Pauli basis, in which $Y=-i\sigma_y$). Recall
that $d=2^N$ is the dimension of the Hilbert space for $N$ qubits.

In order to compare the process  $\mathcal{E}$ with a desired unitary rotation $U$ [i.e.\ with the map $\mathcal{U}(\rho^{\rm in})=U\rho^{\rm in}U^\dagger$], let us formally apply the inverse unitary $U^{-1}=U^\dagger$ after the process $\mathcal{E}$. The resulting composed process
    \be
    \tilde{\mathcal{E}} = \mathcal{U}^{-1} \circ \mathcal{E}
    \label{tilde-E}\ee
characterizes the error: if $\mathcal{E}$ is close to the desired $\mathcal{U}$, then $\tilde{\mathcal{E}}$ is close to the identity (memory) operation.
The process matrix $\tilde{\chi}$ of $\tilde{\mathcal{E}}$ in the Pauli basis is what we call in this paper the process matrix in the Pauli-error basis.

The process matrix $\tilde{\chi}$ obviously satisfies the relation
    \be
        \sum_{\alpha,\beta }\tilde\chi_{\alpha \beta} {\cal P}_{\alpha}\rho^{\rm in} {\cal P}_{\beta}^{\dagger}=
       U^{-1}\left( \sum_{\alpha,\beta }\chi_{\alpha \beta}{\cal P}_{\alpha}\rho^{\rm in} {\cal P}_{\beta}^{\dagger}\right) U,
    \ee
which can be rewritten as
    \be
        \sum_{\alpha,\beta }\tilde\chi_{\alpha \beta} (U{\cal P}_{\alpha}) \rho^{\rm in} (U{\cal P}_{\beta})^{\dagger}=
        \sum_{\alpha,\beta }\chi_{\alpha \beta}{\cal P}_{\alpha}\rho^{\rm in} {\cal P}_{\beta}^{\dagger}.
    \ee
Therefore the error matrix $\tilde\chi$ is formally the process matrix of the original map $\mathcal{E}$, expressed in the operator basis
    \be
    E_\alpha = U{\cal P}_\alpha.
    \ee
This is the Pauli-error basis used in our paper. (Another obvious way to define the error basis is to use $E_\alpha={\cal P}_\alpha U$ \cite{Korotkov-13}; however, we do not use this second definition here.) The Pauli-error basis matrices $E_\alpha$ have the same normalization as the Pauli matrices,
    \be
    \langle E_\alpha | E_\beta \rangle = {\rm Tr} (E^\dagger_\alpha E_\beta)=d\, \delta_{\alpha\beta}.
    \ee
    The matrices $\chi$ and $\tilde\chi$ (in the Pauli and Pauli-error bases) are related via unitary transformation,
    \be
     \tilde\chi=V\chi V^\dagger, \,\,\, V_{\alpha \beta}={\rm Tr}({\cal P}_\alpha^\dagger U^\dagger {\cal P}_\beta)/d.
    \ee

The matrix $\tilde\chi$ has a number of convenient properties
\cite{Korotkov-13}. It has only one large element, which is at the
upper left corner and corresponds to the process fidelity,
$\tilde\chi_{\cal II}=F_\chi=F(\chi, \chi_{\rm ideal})$. All other
non-zero elements of $\tilde\chi$ describe imperfections. In
particular, the imaginary elements in the left column (or upper row)
characterize unitary imperfections (assuming the standard
non-modified Pauli basis), other off-diagonal elements are due to
decoherence, and the diagonal elements correspond to the error
probabilities in the Pauli-twirling approximation.

\section{SVD basis}
\label{AppendixSVD}

The SVD basis used in this paper is introduced following Ref.\
\cite{KosutSVD}. Let us start with the so-called natural basis for
$d\times d$ matrices, which consists of matrices $E^{\rm
nat}_\alpha$, having one element equal to one, while other elements
are zero. The numbering corresponds to the vectorized
form obtained by stacking the columns: for $\alpha = (d-1)i+j$ the matrix is
$(E^{\rm nat}_\alpha )_{lk}=\delta_{il}\delta_{jk}$. For a desired
unitary rotation $U$, the process matrix $\chi^{\rm nat}$ in the
natural basis can be obtained by expanding $U$ in the natural basis,
$U=\sum_\alpha u_\alpha E^{\rm nat}_\alpha$, and then constructing
the outer product,
    \be
    \chi^{\rm nat}_{\alpha\beta}= u_\alpha u_\beta^*.
    \ee
For example, for the ideal CZ gate the components $u_\alpha$ are $(1 ,0 ,0 ,0 ,0 ,1 ,0 ,0 ,0 ,0 ,1 ,0 ,0 ,0 ,0 ,-1)$, and $\chi^{\rm nat}$ has 16 non-zero elements, equal to $\pm 1$.  Note that $\chi^{\rm nat}$ is a rank-1 matrix with ${\rm Tr}(\chi^{\rm nat})=\sum_\alpha |u_\alpha|^2=d$.

We then apply numerical procedure of the SVD decomposition, which diagonalizes the matrix $\chi^{\rm nat}$ for the desired unitary process,
    \be
\chi^{\rm nat}= V \text{diag}(d,0,\ldots, 0) V^{\dagger},
    \label{AppB-decomp}\ee
where $V$ is a unitary $d^2\times d^2$ matrix and the only non-zero eigenvalue is equal to $d$ because ${\rm Tr}(\chi^{\rm nat})=d$. The columns of thus obtained transformation matrix $V$ are the vectorized forms of thus introduced SVD-basis matrices $E_\alpha^{\rm SVD}$,
 \begin{equation}
 E_\alpha^{\rm SVD} =\sum\limits_{\beta=1}^{d^2}V_{\beta \alpha}\, E_{\beta}^{\rm nat}.
 \label{AppB-ESVD}\end{equation}
Note that the notation $V$ used in Appendix A has a different meaning.

The matrices of the SVD basis introduced via Eqs.\ (\ref{AppB-decomp}) and (\ref{AppB-ESVD}) have the different normalization compared with the Pauli basis,
  \begin{equation}
 {\rm Tr} (E_{\alpha}^{{\rm SVD}\dagger}E_{\beta}^{\rm SVD})=\delta_{\alpha \beta}.
 \end{equation}
Correspondingly, the normalization of the process matrix $\chi^{\rm SVD}$ in the SVD basis is ${\rm Tr} \chi^{\rm SVD}=d$ (for a trace-preserving process). For the ideal unitary process the matrix $\chi^{\rm SVD}$ has one non-zero (top left) element, which is equal to $\sqrt{d}$. For an imperfect realization of the desired unitary operation the top left element is related to the process fidelity as $\chi^{\rm SVD}_{11}=F_\chi d$.

Note that when the numerical SVD procedure (\ref{AppB-decomp}) is applied to $\chi^{\rm nat}$ of ideal CZ and/or Toffoli gates, many (most) of the resulting SVD-basis matrices $E_\alpha^{\rm SVD}$ coincide with the matrices of the natural basis $E_\alpha^{\rm nat}$. Since these matrices contain only one non-zero element, the matrix $\Phi$ in Eq.\ (\ref{Vectorized}) is simpler (has more zero elements) than for the Pauli or Pauli-error basis. (The number of non-zero elements of $\Phi$ in the SVD basis is  crudely twice less for the CZ gate and 4 times less for the Toffoli gate.)  As the result, from the computational point of view it is easier to use the SVD basis than the Pauli-error basis: less memory and less computational time are needed.

\section{Average square of state fidelity}
\label{AppendixFidSquared}

In this subsection we present a detailed  derivation of an explicit
formula for the  squared state fidelity $\overline{F^{2}_{\rm st}}$,
averaged over all pure initial states, for a quantum operation,
represented via Kraus operators. We follow the same steps as in
Ref.\ \cite{Emerson2011}, where a closed-form expression for
$\overline{F^{2}_{\rm st}}$ in terms of the process matrix $\chi$
was presented. Although our approach is not new, we show it here for
completeness.

We begin by writing the quantum operation as $\mathcal{E}=\mathcal{U}\circ  \tilde{\mathcal{E}}$ [see Eq.\ (\ref{tilde-E})], where  $\mathcal{U}$ corresponds
to the ideal (desired) unitary operation, while the map $\tilde{\mathcal{E}}$ accounts for the errors in the actual gate. Let
\begin{equation}
\label{Lambda}
  \tilde{\mathcal{E}} (\rho)=\sum_{n} A_{n} \rho A_{n}^{\dagger}
 \end{equation}
 be the operator-sum representation of $\tilde{\mathcal{E}}$,  where $\{A_{n}\}_{n=1}^{d^{2}}$ are Kraus operators satisfying the trace-preservation
 condition $\sum_{n}A^{\dagger}_{n} A_{n}=\mathbb{I} $. The Kraus operators can be easily obtained from the process matrix $\chi_{\alpha
 \beta}$ describing the operation $\mathcal{E}$.  Note that by diagonalizing $\chi$, i.e., $\chi= V D V^{\dagger}$,  where V is
 unitary and $ D=\text{diag}(\lambda_{1}, \lambda_{2},\ldots) $ with $\lambda_{n}\geq 0$,  we can
 express the Kraus operators in Eq.~(\ref{Lambda}) as $A_{n}=\sqrt{\lambda_{n}}\, U^{\dagger} \sum_{\alpha}E_{\alpha} V_{\alpha n}$, where $U$ is the desired unitary.

Now, the state fidelity $F_{\phi}$ (assuming a pure initial state
$|\phi\rangle$)
   can be written in terms of $\{A_{n}\}$ as follows:
 \begin{equation}
F_{\phi} \equiv \bra{\phi}  \tilde{\mathcal{E}} (\phi)\ket{\phi}=
\sum_{n}\bra{\phi}A_{n}\ket{\phi}\bra{\phi}A_{n}^{\dagger}\ket{\phi}.
 \end{equation}
Notice that by using the identity $ \tr(A\otimes B)=\tr(A)\tr(B)$,
one can rewrite the above expression for $F_{\phi}$ as
 \begin{equation}
  F_{\phi}= \sum_{n} \tr{[(A_{n}\otimes A_{n}^{\dagger})( \ket{\phi}\bra{\phi}^{\otimes 2})]},
  \end{equation}
where the notation $\ket{\phi} \bra{\phi}^{\otimes k} \equiv
\underbrace{\ket{\phi}\bra{\phi}\otimes \ket{\phi}\bra{\phi} \ldots
\otimes\ket{\phi}\bra{\phi}}_{k}$ means that the state is copied in
$k$ identical Hilbert spaces. Similarly, one can express the squared
state fidelity as
 \begin{eqnarray}
  \label{fsquaredS4}
  F_{\phi}^{2}&=& \sum_{n,m}\bra{\phi}A_{n}\ket{\phi}\bra{\phi}A_{n}^{\dagger}\ket{\phi}
  \bra{\phi}A_{m}\ket{\phi}\bra{\phi}A_{m}^{\dagger}\ket{\phi}\notag \\
  &=&\sum_{n,m}\tr\big[ (A_{n}\otimes A_{n}^{\dagger}\otimes A_{m}\otimes A_{m}^{\dagger} )
 ( \ket{\phi} \bra{\phi}^{\otimes 4} ) \big]. \qquad
  \end{eqnarray}
In order to compute the average state fidelity $\overline{F_{\rm
st}}=\int F_{\phi}\, d\phi$, the average square of the state
fidelity $\overline{F^{2}_{\rm st}}=\int  F_{\phi}^{2} \, d\phi$,
and higher powers of $F_{\rm st}$ (we assume the normalized integration over the initial pure states, $\int  d\phi=1$), one can use the following
result~\cite{Poulin2004}
  \begin{equation}
  \label{HilbertIntegrals}
  \int \ket{\phi} \bra{\phi}^{\otimes k}  d\phi =
  \frac{1}{ {\binom{k+d-1}{d-1}}} \, \Pi_{k},  \quad
  \Pi_{k}\equiv \frac{1}{k!}\sum_{\sigma \in S_{k}} P_{\sigma}.
  \end{equation}
Here $\sigma$ is an element of the permutation group $S_{k}$ (the
$k!$ permutations of $k$ objects) and the operator $P_{\sigma}$ is
the representation of $\sigma$ in $\mathcal{H}^{\otimes
k}=\underbrace{\mathcal{H}\otimes \ldots \mathcal{H}}_{k}$, i.e.,
  \begin{equation}
  P_{\sigma}(\ket{\phi_{1}}\otimes\ket{\phi_{2}}\ldots \otimes\ket{\phi_{k}})=\ket{\phi_{\sigma(1)}}\otimes \ket{\phi_{\sigma(2)}} \ldots
  \otimes\ket{\phi_{\sigma(k)}}.
  \label{AppC6}\end{equation}
(The operator $P_\sigma$ acts on the wavefunction of $kN$ qubits by permuting $k$ blocks, each containing $N$ qubits.)

In view of the above discussion, we see that the $k$th moment
$\overline{F^{k}_{\rm st}}\equiv \int  F_{\phi}^{k}\, d\phi$ can be
expressed
 as a sum of $(2k)!$ terms corresponding to the elements in
 $S_{2k}$ [note that $k$ in Eqs.\ (\ref{HilbertIntegrals}) and (\ref{AppC6}) is now replaced with $2k$],
\begin{equation}
\label{kth-moment} \overline{F^{k}_{\rm st}}=\frac{\displaystyle \sum_{n_{1}\ldots n_{k}} \sum_{\sigma \in
S_{2k}} \tr [( A_{n_{1}}\otimes
A^{\dagger}_{n_{1}}\otimes \ldots
 A_{n_{k}}\otimes
 A^{\dagger}_{n_{k}}) P_{\sigma}]}{\binom{2k+d-1}{d-1}{(2k)!}}.
 \end{equation}
For example, the average state fidelity $\overline{F_{\rm st}}$ is
determined by the sum over $S_{2}$,
 \begin{eqnarray}
&&  \tr(A_{n} \otimes A_{n}^{\dagger} \ \Pi_{2})= \frac{1}{2}\sum_{\sigma \in S_{2}} \tr(A_{n} \otimes A_{n}^{\dagger} P_{\sigma})\notag \\
&& \hspace{1cm}  =\frac{1}{2}\sum_{\sigma \in
S_{2}}\sum_{i_{1},i_{2}}\bra{i_{1},i_{2}}A_{n}\otimes
A_{n}^{\dagger} \ket{\sigma(i_{1}),
  \sigma(i_{2})}\notag \qquad \\
&& \hspace{1cm}  =\frac{1}{2}(
\underbrace{\tr(A_{n})\tr(A_{n}^{\dagger})}_{\textrm{identity}})+
  \underbrace{\tr(A_{n}A_{n}^{\dagger})}_{\textrm{transposition}}),
  \end{eqnarray}
   which yields the well-known result~\cite{NielsenAveGateFidelity}
  \begin{equation}
\overline{F_{\rm st}}=\frac{1}{d(d+1)}\left( \sum_{n}|\tr(A_{n})|^2+
d\right).
  \end{equation}

  In order to express $\overline{F^{2}_{\rm st}}$ in terms of Kraus operators, it is convenient to write each element of the group $S_{4}$ as a
  product of disjoint cycles.  The 24 elements of the permutation groups $S_{4}$ can be grouped as follows (we use the so-called cycle notation for permutations):

$\bullet$ Identity (1 element): (1)(2)(3)(4) (this notation means that no change of position occurs for all numbers in the sequence 1234);

$\bullet$ Transpositions (6 elements): (12), (13), (14), (23), (24), and (34) (this notations means that only two specified numbers in the sequence are exchanged);

$\bullet$ 3-cycles (8 elements):  (123), (132), (124), (142), (134), (143), (234), and (243) [here the notation (123) means the permutation 1$\rightarrow$2$\rightarrow$3$\rightarrow$1, while the remaining number does not change];

$\bullet$ Products of transpositions (3 elements):  (12)(34), (13)(24), and (14)(23) (two pairs of numbers exchange);

$\bullet$ 4-cycles (6 elements): (1234), (1243), (1324), (1342), (1423), and (1432) [here (1234) means the permutation 1$\rightarrow$2$\rightarrow$3$\rightarrow$4$\rightarrow$1].

This classification simplifies keeping track of the terms $N_{\sigma}\equiv \sum_{n,m}\tr\big[ \big( A_{n}\otimes A_{n}^{\dagger}\otimes
A_{m}\otimes A_{m}^{\dagger} \big) P_{\sigma}\big]$ in Eq.~(\ref{kth-moment}).  The corresponding contributions to the sum $\sum_{\sigma \in
S_{4}} N_{\sigma}$ are the following:
\begin{align*}
&\text{Identity:}\notag \\
& \big(\sum_{n}|\tr(A_{n}\big)|^{2})^{2}. \notag \\
&\text{Transpositions:} \notag \\
& 2d \sum_{n}|\tr(A_{n})|^2+ 2\sum_{n,m}\tr(A_{n}A_{m}^{\dagger})\tr(A_{n}^{\dagger})\tr(A_{m})\notag \\
& + \sum_{n,m}(\tr(A_{n}A_{m})\tr(A_{n}^{\dagger})\tr(A_{m}^{\dagger})+h.c).\notag \\
&\text{3-cycles:}         \notag \notag \\
& 4 \sum_{n}|\tr(A_{n})|^2 +2\sum_{n,m}(\tr(A_{n}A_{n}^{\dagger}A_{m})\tr(A_{m}^{\dagger})+h.c).\notag \\
&\text{Products of transpositions:} \notag \\
&d^2+\sum_{n,m}(|\tr(A_{n}A_{m})|^{2}+ |\tr(A_{n}A_{m}^{\dagger})|^{2}). \notag \\
&\text{4-cycles:}  \notag \\
& 3d +\sum_{n,m}\tr(A_{n}A_{n}^{\dagger}A_{m}A_{m}^{\dagger})+2\sum_{n,m}\tr(A_{n}A_{m}A_{n}^{\dagger} A_{m}^{\dagger}).
\end{align*}
  (We used the trace-preservation condition $\sum_{n} A_{n}^{\dagger}A_{n}=\mathbb{I}$).  Substituting the above terms in
 Eq.~(\ref{kth-moment})  (with $k=2$),  we finally obtain the average square of the state fidelity,
\begin{eqnarray}
\label{fidsquared}
&& \hspace{-0.5cm} \overline{F^{2}_{\rm st}}=\frac{1}{d(d+1)(d+2)(d+3)}\Big(d^{2}+3d
 \nonumber\\
&& +2(d+2)\sum_{n}|\tr(A_{n})|^2+\big( \sum_{n}|\tr(A_{n})|^2\big)^2
 \nonumber\\
&& +\sum_{n,m}\big( |\tr(A_{n}A_{m})|^2+ |\tr(A_{n}A_{m}^{\dagger})|^2\big)
  \nonumber \\
&& + 2\sum_{n,m}\tr(A_{n}A_{m} A_{n}^{\dagger}A_{m}^{\dagger})
+\sum_{n,m}\tr(A_{n}A_{n}^{\dagger}A_{m}A_{m}^{\dagger}) \nonumber\\
&& +2\sum_{n,m} \tr(
A_{n}A_{m}^{\dagger}) \tr(A_{n}^{\dagger})\tr(A_{m})
\nonumber \\
&&+2 \sum_{n,m} {\rm Re}[ \tr(A_{n}A_{m})\tr(A_{n}^{\dagger})\tr(A_{m}^{\dagger})]
 \nonumber \\
&& +4\sum_{n,m} {\rm Re} [\tr(A_{n}A_{n}^{\dagger}A_{m}^{\dagger})\tr(A_{m})
 ] \Big).
\end{eqnarray}
This is the formula we used in this paper to calculate $\overline{F^{2}_{\rm st}}$.



\begin{thebibliography}{60}
\bibitem{s-sparse}
Roughly speaking, a matrix is $s$-compressible if it can be well approximated by an $s$-sparse matrix.
\bibitem{NormDef}
 For a vector $x\in C^m$, the $\ell_{1}$ and $\ell_{2}$ norms are defined as  $|| x ||_{\ell_2} = \sqrt{x^\dagger x} =
 \sqrt{\sum_{i=1}^m|x_i|^2}$ and
$|| x ||_{\ell_1} = \sum_{i=1}^m |x_i|$.
\end{thebibliography}
\end{document}